\documentclass[twocolumn,times]{aastex631}

\usepackage{natbib}
\usepackage{epsfig,amsmath,amssymb}
\usepackage{graphicx}
\usepackage{xcolor}
\usepackage{bm}

\def\CII{[\ion{C}{2}]}
\def\kms{km\,s$^{-1}$}
\def\msunyr{$M_\odot$ yr$^{-1}$}
\def\lya{Ly$\alpha$}
\def\HI{\ion{H}{1}}
\def\zCII{$z_{\rm [CII]}$}
\def \snls{S/N$_{\rm LS}$}

\graphicspath{{./Figs/}}

\begin{document}

\title{A [C\,\textsc{ii}] 158$\bm \mu$m Survey of DLA galaxies at $\bm{z \sim 4}$: Observations of Dense and Metal-Enriched Neutral Gas within the Circumgalactic Medium of Star-Forming Galaxies}
\shorttitle{A [C\,\textsc{ii}] Survey of DLA galaxies at $z \sim 4$}
 \shortauthors{Neeleman et al.}

\correspondingauthor{Marcel Neeleman}
\email{mneeleman@nrao.edu}

\author[0000-0002-9838-8191]{Marcel Neeleman}
\affiliation{National Radio Astronomy Observatory, 520 Edgemont Road, Charlottesville, VA, 22903, USA}

\author[0000-0002-9757-7206]{Nissim Kanekar}
\affiliation{National Centre for Radio Astrophysics, Tata Institute of Fundamental Research, Pune University, Pune 411007, India}

\author[0000-0002-7738-6875]{J. Xavier Prochaska}
\affiliation{Department of Astronomy \& Astrophysics, UCO/Lick Observatory, University of California, 1156 High Street, Santa Cruz, CA 95064, USA}
\affiliation{Kavli Institute for the Physics and Mathematics of the Universe (Kavli IPMU), 5-1-5 Kashiwanoha, Kashiwa, 277-8583, Japan}

\author[0000-0002-9946-4731]{Marc A. Rafelski}
\affiliation{Space Telescope Science Institute, 3700 San Martin Drive, Baltimore, MD 21218, USA}
\affiliation{The William H. Miller III Department of Physics \& Astronomy, Johns Hopkins University, Baltimore, MD 21218, USA}

\author[0009-0007-5296-4046]{Lordrick A. Kahinga}
\affiliation{Department of Astronomy \& Astrophysics, UCO/Lick Observatory, University of California, 1156 High Street, Santa Cruz, CA 95064, USA}
\affiliation{Maria Mitchell Observatory, Nantucket, MA, 02554, USA}
\affiliation{Department of Physics, College of Natural and Mathematical Sciences, University of Dodoma, 41218 Iyumbu, Dodoma, Tanzania}

\begin{abstract}
We present a survey undertaken with the Atacama Large Millimeter/submillimeter Array (ALMA) to study the galaxies associated with a representative sample of 16 damped \lya\ absorbers (DLAs) at $z\approx4.1-4.5$, using the \CII~158$\mu$m (\CII) line. We detect seven \CII-emitting galaxies in the fields of 5 DLAs, all of which have absorption metallicity [M/H]~$> -1.5$. We find that the detectability of these \HI-selected galaxies with ALMA is a strong function of DLA metallicity, with a detection rate of $71^{+11}_{-20}$\% for DLAs with [M/H]~$> -1.5$ and $0^{+18}$\% for DLAs with [M/H]~$<-1.5$. The identified DLA galaxies have far-infrared properties similar to those of typical star-forming galaxies at $z \sim 4$, with estimated obscured star-formation rates ranging from $\lesssim 6$~\msunyr\ to  110~\msunyr. High-metallicity DLAs therefore provide an efficient way to identify and study samples of high-redshift, star-forming galaxies without preselecting the galaxies by their emission properties. The agreement between the velocities of the metal absorption lines of the DLA and the \CII\ emission line of the DLA galaxy indicates that the metals within the DLA originated in the galaxy. With observed impact parameters between 14 and 59 kpc, this indicates that star-forming galaxies at $z \sim 4$ have a substantial reservoir of dense, cold neutral gas within their circumgalactic medium that has been enriched with metals from the galaxy. 
\end{abstract}


\section{Introduction}
\label{sec:intro}

Ever since the realization that quasar spectra contained \lya\ absorption features blueshifted with respect to the quasar \lya\ emission \citep{Greenstein1967,Bahcall1968,Burbidge1968}, astronomers have tried to search for the sources responsible for this absorption. The strongest \lya\ absorbers have \HI\ column densities of $N_{\rm HI} \geq 2 \times 10^{20}$ cm$^{-2}$ and are known as damped \lya\ absorbers \citep[DLAs;][]{Wolfe2005}. Unlike lower \HI\ column density systems that often arise from the trace
neutral atoms located within the intergalactic medium \citep[e.g.,][]{Rauch1998}, DLAs are expected to arise from gas located either within galaxies or in the immediate surroundings of galaxies \citep{Wolfe1986}. This expectation is based on several independent observations and simulations. First, local observations of \HI\ have shown that column densities of $N_{\rm HI} \geq 2 \times 10^{20}$ cm$^{-2}$ arise almost exclusively in the disks of star-forming galaxies \citep[e.g.,][]{Ryan-Weber2003,Zwaan2005}. Second, cross-correlation studies between DLAs and the \lya\ forest have shown that DLAs cluster in a manner similar to galaxies \citep{Font-Ribera2012,Perez-Rafols2018a}. Finally, cosmological simulations require a connection between DLAs and star-formation in order to correctly match the number density of galaxies at high redshifts \citep[e.g.,][]{Pontzen2008,Altay2011,Bird2014,Diemer2019,DiGioia2020}.

However, directly studying the galaxies that are expected to be near DLAs, which we will label in this paper as DLA galaxies or \HI-selected galaxies\footnote{This is an observational definition describing the selection technique that was used to find the galaxy, in the same way as Lyman Break galaxies and \lya-emitting galaxies have been defined. As such, this does not describe any unique properties of the actual galaxy or galaxy population. It is also distinct from the term ``DLA host galaxy'', which we do not use in this paper, because the term ``host'' can be ambiguous for some DLAs where the gas could arise from multiple, distinct origins (see Section \ref{sec:dlagal})}, remains challenging \citep[e.g.,][]{Fumagalli2014,Krogager2017,Fynbo2023}. This is primarily due to the expected faintness of typical DLA galaxies at optical and near-IR wavelengths, combined with the brightness of the background QSO. Simulations predict that DLA galaxies have stellar masses of $\lesssim 10^8 M_\odot$ \citep{Bird2014}, which pushes the detection limits of current observational facilities. Some observational studies have leveraged the metallicity of the DLA to increase the detection rate of DLA galaxies \citep[e.g.][]{Fynbo2010,Krogager2017}. This metallicity-based selection technique relies on the observational result that stellar mass and metallicity are correlated \citep{Tremonti2004,Erb2006,Mannucci2009,Maiolino2019}. Although the DLA metallicity may not be equal to the metallicity of the DLA galaxy, the success of this approach suggests that some connection exists between the mass of the DLA galaxy and the metallicity of the DLA. This is further supported by the observed correlation between the DLA metallicity and the equivalent width of the metal absorption lines of the DLA \citep{Ledoux2006,Prochaska2008,Moller2013,Neeleman2013}. The latter has been used as a rough proxy for the mass of the DLA galaxy \citep[e.g.,][]{Christensen2014}. 

Using the metallicity of the DLA as a preselection technique, several studies have successfully used the Atacama Large Millimeter/sub-millimeter Array (ALMA), to find and study DLA galaxies \citep[e.g.,][]{Neeleman2016,Neeleman2017,Neeleman2018,Neeleman2019,Kanekar2018,Kanekar2020,Kaur2022}. At $z \lesssim 2$, these studies use low-J and mid-J CO emission lines to identify and characterize DLA galaxies. \citet{Kanekar2020} report a success rate of $>$40~\% in detecting DLA galaxies at $z \sim 2$ when targeting DLAs with a metallicity [M/H]~$> -0.72$, and the galaxy detection rate is $>$70~\% at $z < 1$ \citep{Kanekar2018}. \citet{Kaur2022} used the NOrthern Extended Millimeter Array (NOEMA) and ALMA to show a strong dichotomy in $z \sim 2$ DLA galaxy detection rates when targeting DLAs with [M/H] $< -0.3$ and those above this threshold. The large molecular masses for the DLA galaxies associated with DLAs with a metallicity $> -0.3$ further support the scenario that the highest-metallicity DLAs are associated with massive galaxies. 

Recent follow-up observations of some of these systems using the Keck Cosmic Web Imager on the Keck~II telescope \citep{Oyarzun2024} have revealed two interesting features. First, the CO-emitting DLA galaxies are not detected in \lya\ emission. This is similar to the scenario at lower redshifts ($z \sim 0.3$) whereby the optical observations and millimeter observations provide a complementary view of the fields surrounding DLAs \citep[e.g.,][]{Peroux2019}. Second, no \lya\ emitters are seen closer to the DLA than the CO-emitting galaxy. This leads to the conclusion that high-metallicity DLAs predominantly probe \HI\ in the outskirts of the DLA galaxy observed in CO emission, either within an extended disk or within the circumgalactic medium (CGM) of the galaxy.

At $z \gtrsim 3$, recent studies using the Multi Unit Spectroscopic Explorer on the Very Large Telescope have targeted DLAs with a range of metallicities \citep{Mackenzie2019,Lofthouse2023}. These studies find a high ($\approx$80~\%) detection rate of galaxies associated with the DLA. However, most of these DLA galaxies have large impact parameters, $\gtrsim 100$~kpc. The DLA sightline is thus probing gas that is outside the virial radius of the galaxies, making it unlikely that these galaxies are directly responsible for the DLA. These studies hence argue that the identified DLA galaxies found are part of overdensities traced by the DLA. In \citet{Mackenzie2019}, the two galaxy--DLA pairs whose impact parameters are smaller than the virial radii have DLAs with a relatively high metallicity of [M/H] $= -1.1$. This again illustrates the strong dependence on metallicity in detecting galaxies for which the DLA sightline probes gas within the virial radius of the galaxy. 

At $z \gtrsim 4$, ALMA has enabled the detection of DLA galaxies using the $157.7\mu$m fine-structure line of singly ionized carbon \citep[the \CII 158$\mu$m line; hereafter, \CII;][]{Neeleman2017,Neeleman2019}. These studies targeted DLAs with [M/H] $> -1.3$, finding a DLA galaxy detection rate of $\approx 80$~\% with impact parameters up to $\approx 40$~kpc. Although the DLA sightlines are within the virial radii, a smaller satellite galaxy might also be responsible for the DLA, as suggested by simulations \citep{Rahmati2014}. However, \citet{Neeleman2019} also find that the velocity difference between emission and absorption is $\lesssim$100~\kms, which runs counter to the scenario of an orbiting satellite galaxy. The limited metallicity range of the DLAs precluded \citet{Neeleman2019} from making any definitive statement on the efficacy of the metallicity selection, and further prevented any general statement on the ability of ALMA to detect DLA galaxies at $z > 4$.

In this paper, we present ALMA observations of a sample of 16 DLAs selected solely based on their redshift, the ability to be observed by ALMA, and the availability of an accurate metallicity measurement. This sample is representative of the general DLA population at $z \sim 4$ in both the \HI\ column density and metallicity distributions. Therefore, this study should provide a definitive answer on the ability to detect $z \gtrsim 4$ DLA galaxies in \CII\ emission with ALMA. In Section \ref{sec:sample}, we discuss the sample selection. In Section \ref{sec:observations}, we outline the observations and the data reduction procedure. The methods used to search for emission from the DLA fields are described in Section \ref{sec:methods}, and the search results presented in Section~\ref{sec:analysis}. The identified emission lines and implications for DLA galaxies are discussed in Section~\ref{sec:discussion}, and the paper is summarized in Section~\ref{sec:summary}. Throughout the paper, we will use a standard, flat $\Lambda$ cold dark matter cosmology with $H_0 = 70$ \kms\ Mpc$^{-1}$ and $\Omega_{\rm M} = 0.3$. For reference, in this cosmology, 1$''$ corresponds to 6.8~physical kpc at $z = 4.2$. Throughout the paper, we report physical distances, unless otherwise indicated.

\begin{deluxetable*}{llrcccr}
\tablecaption{DLA Properties
\label{tab:dlaprop}}
\tablehead{
\colhead{QSO} &
\colhead{R.A.} &
\colhead{Decl.} &
\colhead{$z_{\rm abs}$} &
\colhead{log ($N_{\rm HI}/{\rm cm}^{-2}$} &
\colhead{[M/H]} &
\colhead{$\Delta v_{90}$}\\
\colhead{} &
\colhead{(J2000)} &
\colhead{(J2000)} &
\colhead{} &
\colhead{} &
\colhead{} &
\colhead{(km s$^{-1}$)}
}
\startdata
J0307-4945 & 03:07:22.89 & $-$49:45:48.3 & 4.4679 & 20.67$\pm$0.09 & -1.45$\pm$0.12 & 192\\
J0817+1351 & 08:17:40.53 & +13:51:34.6 & 4.2584 & 21.30$\pm$0.15 & -1.15$\pm$0.15 &  87\\
J0824+1302 & 08:24:54.01 & +13:02:17.0 & 4.4720 & 20.30$\pm$0.10 & -1.97$\pm$0.12 &  70\\
J0834+2140 & 08:34:29.44 & +21:40:24.7 & 4.3900 & 21.00$\pm$0.20 & -1.30$\pm$0.20 & 290\\
J0834+2140 & 08:34:29.44 & +21:40:24.7 & 4.4610 & 20.30$\pm$0.15 & -1.85$\pm$0.15 &  70\\
BR0951-04 & 09:53:55.75 & $-$05:04:19.0 & 4.2029 & 20.40$\pm$0.10 & -2.57$\pm$0.10 &  28\\
J1054+1633 & 10:54:45.43 & +16:33:37.4 & 4.1346 & 21.00$\pm$0.10 & -0.70$\pm$0.11 & 200\\
J1100+1122 & 11:00:45.23 & +11:22:39.1 & 4.3947 & 21.74$\pm$0.10 & -1.68$\pm$0.21 & 142\\
J1101+0531 & 11:01:34.36 & +05:31:33.9 & 4.3446 & 21.30$\pm$0.10 & -1.07$\pm$0.12 &  60\\
J1132+1209 & 11:32:46.50 & +12:09:01.7 & 4.3802 & 21.20$\pm$0.20 & -2.57$\pm$0.17 & 100\\
BR1202-07 & 12:05:23.14 & $-$07:42:32.8 & 4.3829 & 20.60$\pm$0.14 & -1.75$\pm$0.14 & 172\\
PSS1443+27 & 14:43:31.17 & +27:24:36.8 & 4.2241 & 21.00$\pm$0.10 & -0.95$\pm$0.20 &  90\\
J1607+1604 & 16:07:34.22 & +16:04:17.4 & 4.4741 & 20.30$\pm$0.15 & -1.71$\pm$0.15 &  42\\
J1626+2751 & 16:26:26.50 & +27:51:32.5 & 4.3110 & 21.34$\pm$0.15 & -1.34$\pm$0.17 & 120\\
\nodata & \nodata & \nodata & 4.4975 & 21.39$\pm$0.15 & -2.46$\pm$0.16 & 250\\
PSS2241+1352 & 22:41:47.76 & +13:52:02.7 & 4.2833 & 21.15$\pm$0.10 & -1.72$\pm$0.11 &  70
\enddata
\tablecomments{[M/H], $z_{\rm abs}$ and log ($N_{\rm HI}/{\rm cm}^{-2}$ are from \citet{Rafelski2012} and references therein. The $\Delta v_{90}$ values are from \citet{Neeleman2013}.}
\end{deluxetable*}

\begin{figure}[!tbh]
\includegraphics[width=0.48\textwidth]{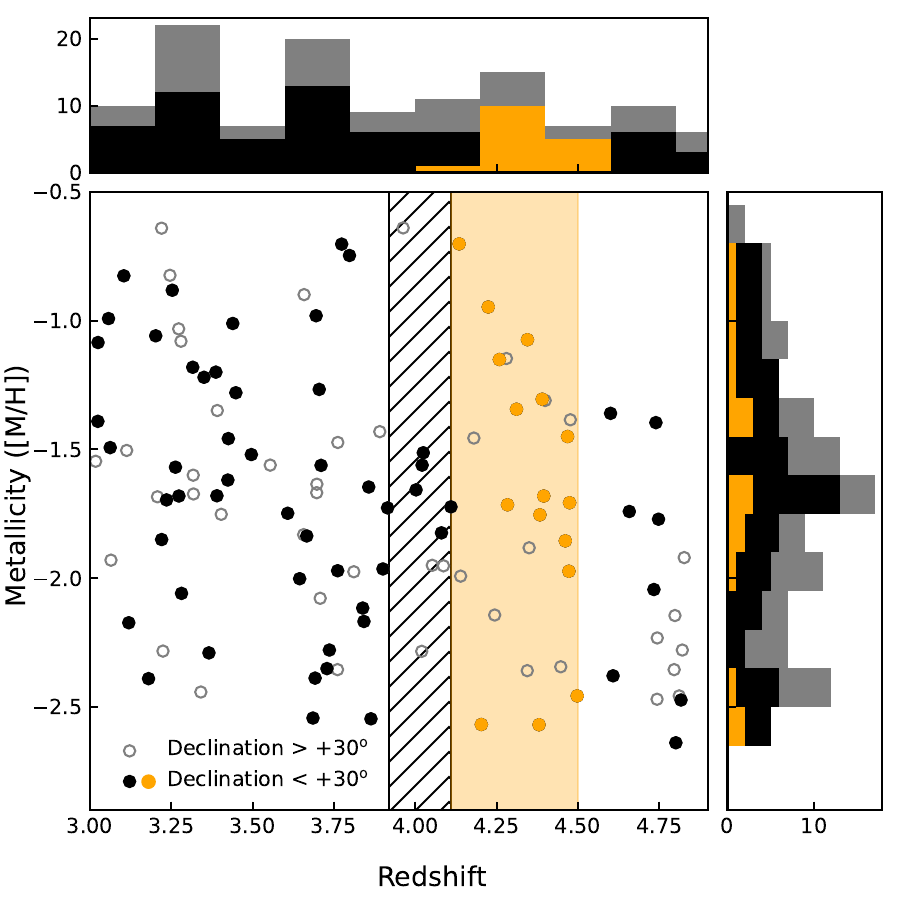}
\caption{Redshift and metallicity distributions of the sample of DLAs targeted with ALMA in this study. The parent DLA population described in \citet{Rafelski2014} has been subdivided into those outside the observing range of ALMA (i.e., declinations $<$+30\degr; open gray circles), and those within the observing range (solid circles). The orange circles are the DLAs of our sample and constitute all of the DLAs within the redshift range $z = 4.11 - 4.5$ (light orange-shaded region). The hatched region marks the region in redshift space that is unobservable with ALMA in \CII, because the redshifted \CII\ emission line for galaxies in this redshift range falls in the frequency range between ALMA Bands~7 and 8. 
\label{fig:zmh}}
\end{figure}

\section{The DLA Sample}
\label{sec:sample}

Our aim in constructing the sample is to introduce as little selection bias as possible. For this, we start with the parent sample of DLAs described in \citet{Rafelski2012,Rafelski2014}. This parent sample consists of 258 DLAs that were observed with either medium- or high-resolution ($R > 7000$) optical spectrographs and that were not selected based on any \textit{a priori} knowledge of the metal content of the DLA. Although the parent sample is biased toward higher \HI\ column density systems at the DLA threshold compared to larger DLA samples based on \HI\ column density alone \citep[e.g., from the Sloan Digital Sky Survey; ][]{Prochaska2005,Noterdaeme2012}, this is not likely to introduce a significant bias in the DLA properties, because \HI\ column density is not strongly correlated with any other DLA property \citep[e.g.,][]{Neeleman2013}, with the possible exception of the impact parameter between the DLA sightline and the DLA galaxy \citep{Noterdaeme2014,Rahmati2014}. The high resolution spectra allow for accurate metallicity measurements for all the DLAs in the sample, and also provide information on the DLA kinematics through the absorption line profiles \citep{Prochaska1997,Neeleman2013}. 

From this parent sample, we select all DLA fields that are observable with ALMA (i.e., declination $<$+30\degr). We further consider only those DLAs that lie within the redshift range $4.11 < z < 4.5$. The lower redshift limit is set by our goal to target \CII\ emission from galaxies associated with the DLAs. For $z=4.11$, this corresponds to a sky frequency of $\approx 371.93$~GHz, which is the highest frequency that can be observed with ALMA in Band~7 while still allowing for an $\approx$1~GHz spectral range on either side of the \CII\ emission line. We note that the next ALMA band (band 8) does not start until 385~GHz, corresponding to \zCII~$=3.92$. The upper redshift limit is set by contraints on the sample size based on the availability of ALMA observing time. This results in a sample of 16 DLAs with a median redshift of 4.38 and a median metallicity ([M/H]\footnote{[M/H] is defined as the logarithmic abundance of metals with respect to solar such that [M/H] = 0 corresponds to 12 + log(O/H) = 8.69 \citep{Asplund2009}.}) of $-1.7$. 

The individual details of the DLAs are tabulated in Table~\ref{tab:dlaprop}. The redshift and metallicity distributions of the sample, and of the parent sample of \citet{Rafelski2014} for the redshift range $3 < z < 5$, are shown in Fig.~\ref{fig:zmh}. The metallicity is measured from unsaturated metal absorption lines in the DLA using the apparent optical depth method following the procedure described in \citet{Rafelski2012}. In short, we use unsaturated \ion{S}{2}, \ion{Si}{2}, \ion{Zn}{2} or \ion{Fe}{2} lines to measure the metallicity of the DLA. In case \ion{Fe}{2} lines are used, we apply a 0.3~dex correction, to account for dust depletion and $\alpha$-enhancement. The measurement uncertainties take into account the small systematic uncertainties in using these different tracers. The metallicity histogram shows that the sample of this paper is representative of the metallicity distribution over the larger redshift range, despite the metallicity evolution across this redshift range. We therefore assert that this sample is representative of the DLA population at $z \gtrsim 4$. 

\section{Observations and Data Reduction}
\label{sec:observations}

\begin{deluxetable*}{lcllcllc}
\tablecaption{Journal of the ALMA observations
\label{tab:obs}}
\tablehead{
\colhead{QSO} &
\colhead{$t_{\rm on-source}$} &
\colhead{$\nu_{\rm cont}$} &
\colhead{Beam$_{\rm cont}$} &
\colhead{rms$_{\rm cont}$} &
\colhead{$\nu_{\rm [CII]}$} &
\colhead{Beam$_{\rm [CII]}$} &
\colhead{rms$_{\rm [CII],31MHz}$}\\
\colhead{} &
\colhead{(s)} &
\colhead{(GHz)} &
\colhead{($'' \times ''$)} &
\colhead{($\mu$Jy beam$^{-1}$)} &
\colhead{(GHz)} &
\colhead{($'' \times ''$)} &
\colhead{(mJy beam$^{-1}$)}
}
\startdata
J0307-4945 & 7890 & 341 & $0\farcs95 \times 0\farcs88$ & 17 & 347.578 & $0\farcs94 \times 0\farcs85$ & 0.21\\
J0817+1351 & 2780 & 356 & $0\farcs97 \times 0\farcs86$ & 37 & 361.429 & $0\farcs94 \times 0\farcs84$ & 0.46\\
J0824+1302 & 4480 & 342 & $0\farcs98 \times 0\farcs84$ & 25 & 347.320 & $0\farcs96 \times 0\farcs82$ & 0.29\\
J0834+2140 & 7200 & 345 & $0\farcs70 \times 0\farcs57$ & 19 & 352.604 & $0\farcs69 \times 0\farcs56$ & 0.25\\
J0834+2140 & 5620 & 343 & $0\farcs95 \times 0\farcs91$ & 23 & 348.020 & $0\farcs96 \times 0\farcs88$ & 0.30\\
BR0951-04 & 7800 & 360 & $0\farcs89 \times 0\farcs72$ & 23 & 365.284 & $0\farcs85 \times 0\farcs68$ & 0.38\\
J1054+1633 & 1570 & 363 & $0\farcs98 \times 0\farcs94$ & 57 & 370.143 & $1\farcs06 \times 0\farcs89$ & 1.10\\
J1100+1122 & 11250 & 345 & $1\farcs12 \times 0\farcs86$ & 15 & 352.297 & $1\farcs10 \times 0\farcs85$ & 0.26\\
J1101+0531 & 2750 & 349 & $0\farcs94 \times 0\farcs85$ & 29 & 355.599 & $0\farcs92 \times 0\farcs83$ & 0.40\\
J1132+1209 & 4840 & 348 & $1\farcs09 \times 0\farcs77$ & 23 & 353.247 & $1\farcs06 \times 0\farcs76$ & 0.35\\
BR1202-07 & 4110 & 348 & $0\farcs90 \times 0\farcs65$ & 32 & 353.073 & $0\farcs88 \times 0\farcs64$ & 0.40\\
PSS1443+27 & 7990 & 357 & $0\farcs69 \times 0\farcs48$ & 29 & 363.802 & $0\farcs68 \times 0\farcs47$ & 0.36\\
J1607+1604 & 4720 & 342 & $0\farcs94 \times 0\farcs90$ & 21 & 347.187 & $0\farcs93 \times 0\farcs88$ & 0.27\\
J1626+2751 & 6990 & 353 & $1\farcs14 \times 0\farcs89$ & 26 & 357.849 & $1\farcs11 \times 0\farcs88$ & 0.37\\
\nodata & \nodata & \nodata & \nodata & \nodata & 345.709 & $1\farcs16 \times 0\farcs90$ & 0.29\\
PSS2241+1352 & 4660 & 355 & $0\farcs93 \times 0\farcs77$ & 30 & 359.725 & $0\farcs92 \times 0\farcs75$ & 0.42
\enddata
\tablecomments{$\nu_{\rm [CII]}$ is the expected redshifted frequency of the \CII\ emission line based on the DLA redshift. The two DLAs towards J1626+2751 were observed simultaneously.}
\end{deluxetable*}

The observations were conducted as part of three separate ALMA observing programs: 2015.1.01564.S (PI: Neeleman), 2016.1.00569.S (PI: Neeleman) and 2018.1.01784.S (PI: Prochaska),
between December 2015 and March 2019, using the Band-7 receivers. One of the DLAs of the sample, J0817$+$1351, was discussed in \citet{Neeleman2017}. This field was followed up with higher resolution ALMA \CII\ observations to explore the kinematics of the \CII\ emission, revealing a massive rotating disk galaxy \citep[the ``Wolfe disk'';][]{Neeleman2020}. Also, in \citet{Neeleman2019}, we reported results from the \CII\ observations of three more sources of this sample, J1101$+$0531, J0834$+$2140, and PSS1443$+$27. The observations and data reduction of the remaining 12 sources are presented here. In passing, we note that the \CII\ images of J1054$+$1633 have been shown in \citet{Kaur2021}.

All of our targets were observed with ALMA in a compact configuration with mean maximum baseline of $\approx 300$~m. The correlator was set up such that one spectral window covered the redshifted \CII\ emission line of the DLA target. For programs 2015.1.01564.S and 2016.1.00569.S, this spectral window used the Time Division Mode (TDM), with a bandwidth of 2~GHz and 128 channels, yielding a channel spacing of 15.625~MHz. For program 2018.1.01784.S, this spectral window used the Frequency Division Mode (TDM), with a bandwidth of 1.875~GHz and 240 channels, yielding a channel spacing of 7.8125~MHz. For all three programs, the remaining three spectral windows were set up in TDM to observe the continuum emission in the field, using a bandwidth of 2~GHz, 128~channels and a channel spacing of 15.625~MHz. The only exception to the above setup was for the observations of QSO~J1626+2751 (program 2018.1.01784.S), which has two DLAs along the sightline that were observed simultaneously. Here, the two spectral windows covering the expected redshifted \CII\ line of the two DLAs were set up in FDM, each with a 1.875~GHz bandwidth and 240~channels, while the other two spectral windows were set up in TDM, each with a 2~GHz bandwidth and 128~channels to cover the continuum emission. The total on-source time per target varied between 45~min and 3h. We list details of the observations of individual sources in Table~\ref{tab:obs}. 

The observations were initially processed using the ALMA pipeline \citep{Hunter2023}, which is part of the common astronomy software applications package \citep[CASA;][]{CASA2022}. The observations extended over several ALMA observing cycles, and we hence used different versions of the ALMA pipeline for the initial analysis, depending on the recommendations for each cycle. After the ALMA pipeline data editing and calibration, each data set was analyzed independently (by M.N. and N.K., using similar but independent procedures to partially mitigate the effect of the reduction procedure on the data products). This involved first inspecting the calibrated data products created by the ALMA pipeline.  For several data sets, we manually excised additional baselines or antennas using the data editing functions in CASA. By design, all of our fields contain a quasar at the phase center of the pointing. For quasars bright enough in the continuum (typically, flux densities $\gtrsim$2~mJy~beam$^{-1}$), we performed self-calibration using either the CASA calibration tasks or the independent calibration routine \texttt{GAINCALR} \citep{Chowdhury2022}. In practice, the self-calibration was performed using either two or three rounds of phase-only self-calibration, and, depending on the quasar flux density, one round of amplitude and phase calibration. The resultant antenna-based gains were then applied to the multi-channel visibility data set, before producing the spectral cubes.

The final continuum images and spectral cubes were created using the task \texttt{tclean} in CASA~v6.5.2. To create the continuum images, we used natural weighting, and excluded any spectral ranges that contain a bright emission line within the field. For each source, we then subtracted the continuum emission from the calibrated multi-channel visibilities, to produce a residual visibility data set. To accomplish this we either used the task \texttt{uvcontsub} (M.N) or the task \texttt{uvsub} (N.K.) to subtract out any identified continuum. The residual visibilities were then imaged in channels of 31.25~MHz width ($\approx 30$~\kms, at the DLA redshifts) using natural weighting and either a multi-scale clean deconvolver (M.N) or the standard H\"{o}gbom deconvolver (N.K.). Similar results were obtained via the two deconvolution algorithms. The two independent analyses also ``cleaned'' the emission to different levels, down to twice the root mean square (rms) noise of each spectral  channel (M.N.), and to the rms noise of each channel (N.K.); similar results for the final line flux densities were obtained with the two approaches. The rms sensitivities and synthesized beam sizes for both the continuum images and the spectral cubes are listed in Table~\ref{tab:obs}. 

We note that the varying rms sensitivity between the different fields could result in a bias of finding \CII\ emitters predominantly in the fields with better sensitivity. As we show in Section \ref{sec:analysis}, most \CII\ emitters are actually found in the fields with poorer sensitivity. As a result, the rms sensitivity variability does not significantly affect the conclusions of this paper.

\section{Methods}
\label{sec:methods}

In this section, we describe the two methods that were used to identify emission lines in the ALMA spectral cubes. In subsection \ref{sec:autoline}, we describe two automated line-finding algorithms. These were designed to find emission lines in interferometric data. In subsection \ref{sec:manline}, we describe a manual approach towards finding emission lines around the DLA redshifts. This approach has been used in previous ALMA studies that searched for the galaxies associated with DLAs \citep[e.g.,][]{Neeleman2017,Neeleman2018,Neeleman2019,Fynbo2018,Kanekar2020,Kaur2022}. 

\subsection{Automated line searches}
\label{sec:autoline}

\begin{figure}[!tbh]
\includegraphics[width=0.48\textwidth]{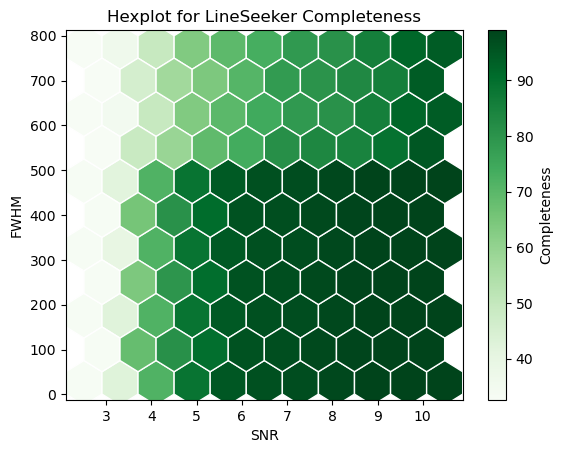}
\includegraphics[width=0.48\textwidth]{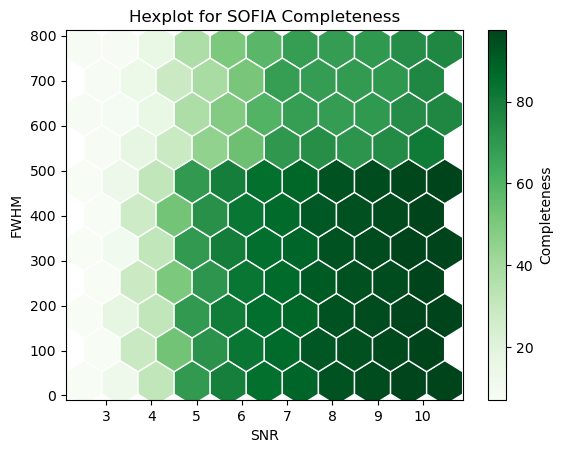}
\caption{Completeness estimate as a function of S/N and FWHM of the emission line for LineSeeker (top panel) and SoFiA (bottom panel). The colors denote the percentage of sources recovered within each bin. The figure shows that LineSeeker is able to recover a higher percentage of sources compared to SoFiA, with completion percentages $> 90$~\% for sources with a FWHM velocity $<$ 500 \kms\ and S/N $>$ 5. 
\label{fig:completeness}}
\end{figure}

\begin{figure}[!tbh]
\includegraphics[width=0.48\textwidth]{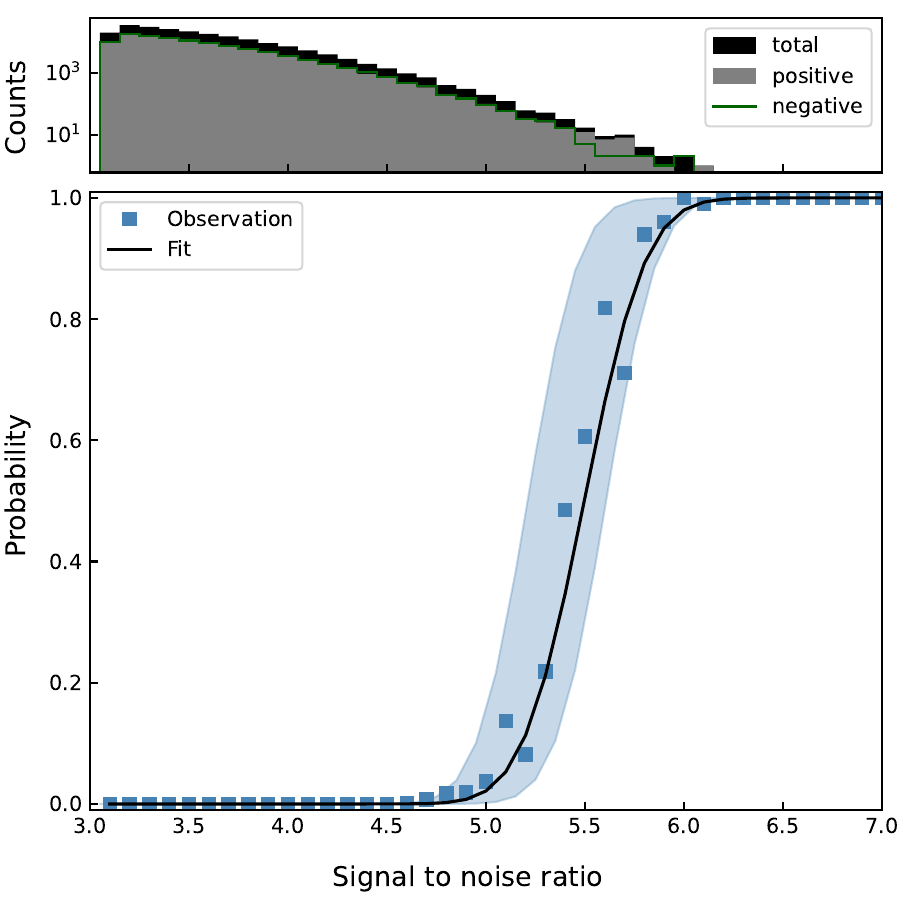}
\caption{Fidelity of the source detection. Top panel shows the total number of positive and negative sources detected within the data cubes binned by S/N. For pure noise, we would expect both the negative and positive number counts to be equal, which is the case below a S/N of 5. The probability for a source to be real is shown in the main panel. For S/N greater than 5.7, the probability that a source is real is greater than 90 \%. This S/N ratio is taken as our detection threshold in the remainder of the paper. 
\label{fig:fidelity}}
\end{figure}

The primary aim of the ALMA observations is to search for \CII\ emission lines from galaxies associated with DLAs at $z \approx 4$. However, the observations cover a much larger frequency range (roughly 7~GHz per target) over which we can search for line emitters. These observations are among the deepest at this frequency range, comparable in depth to the ALMA large program, ASPECS \citep{Walter2016}, which probed the Hubble Ultra Deep Field at $\approx$240~GHz. Besides the searches for line emitters at or near the DLA redshifts, we hence also carried out a ``blind'' search for line emitters in all our spectral cubes. Such a blind search can benefit from an automated approach to find and classify the line candidates for two reasons. First, it allows us to explore potential serendipitous lines within the spectral cubes. Second, it provides a crucial check on the reliability of potential line emitters near the DLA redshift, based on the actual statistical properties of the spectral cube (see Section~\ref{sec:manline}).

For the automated search for line emitters, we tested two search algorithms: LineSeeker \citep{Gonzalez-Lopez2017} and SoFiA \citep{Serra2015,Westmeier2021}. The former is a line search tool designed primarily to search for unresolved emission lines in (sub-)millimeter-wave spectral cubes, whereas the latter was designed to search for extragalactic \HI\ 21\,cm emission signals in wide-field spectral cubes at low (GHz) frequencies. Both tools convolve the data cube in the spectral direction, using either a top-hat or a Gaussian kernel. In addition, SoFiA has the capability to convolve the data cube in the spatial direction to look for extended emission. For a detailed description of the two search tools we refer the reader to \citet{Gonzalez-Lopez2017}, 
 \citet{Serra2015} and \citet{Westmeier2021}.
 
To discern which line search tool is most effective for our study, we injected artificial sources into simulated ALMA data cubes that span the range of beam sizes and sensitivities of our observational data. These artificial sources were point sources with a velocity full width at half maximum (FWHM) in the range $50 - 800$~\kms, and with varying signal-to-noise ratios (S/N), in the range $2 - 10$. These choices were motivated by the typical sizes and velocity widths of the \CII\ emission from main-sequence galaxies at $z\approx4$ from both DLA studies \citep{Neeleman2019} and emission-selected samples \citep[e.g.,][]{Fujimoto2020}. These studies have shown that galaxies have typical \CII\ effective radii of $<3$~kpc and that most of this emission will arise from  within a single beam of our observations, which have a typical FWHM beam size of 5~kpc at the DLA redshifts. We then used both LineSeeker and SoFiA to try and recover the emission lines, i.e., to study the completeness of the search algorithm. The completeness of the two algorithms is shown in Fig.~\ref{fig:completeness}. Although SoFiA employs a more sophisticated search algorithm than LineSeeker, allowing for the detection of resolved emission, its completeness in the case of unresolved sources is slightly worse than LineSeeker. Since we are primarily expecting unresolved emission in our sub-millimeter spectral cubes, we adopted LineSeeker as our main automated search tool. 

We carried out the search for emission lines separately in all four spectral windows of each ALMA data set. For each cube, we convolve the cube in the spectral direction with a Gaussian with a width ranging from 1 to 17 channels. This optimizes our search for sources with FWHM up to 1000~\kms. To estimate the fidelity of a line detection, we compare the number of positive ($n_{\rm pos, \Delta v}$) and negative ($n_{\rm neg, \Delta v}$) sources that are detected at a given \snls\footnote{The S/N definition employed by LineSeeker differs slightly from the S/N calculation adopted elsewhere; we have hence used ``LS'' as the subscript of the S/N for LineSeeker searches (see Section~\ref{sec:manline}).} in the spectrally convolved data cube that yielded the detection \citep[e.g.,][]{Gonzalez-Lopez2019,Decarli2020,Venemans2020}. The probability of an identified spectral feature being real (i.e., the line fidelity) is then defined as: $P_{\rm obs}({\rm S/N_{LS}}) = 1 - n_{\rm neg, \Delta v}({\rm S/N_{LS}}) / n_{\rm pos, \Delta v}({\rm S/N_{LS}})$. This probability estimate is accurate if the noise is Gaussian and all of the negative sources are due to noise fluctuations, but it ignores any systematics affecting the data cube (e.g., due to incorrect continuum subtraction, atmospheric lines, etc.).

Our data sets have similar angular resolutions and sensitivity, and we find little evidence for any source-specific dependence in the line fidelity as a function of \snls. In Fig. \ref{fig:fidelity}, we therefore show the cumulative properties of all data sets. We note that for a given \snls, emission lines with a larger velocity width will have a higher probability of being real \citep[][and Fig. \ref{fig:completeness}]{Gonzalez-Lopez2019}, but this effect is small compared to the stochasticity of each individual \snls\ bin. We therefore perform a fit to the median probability as a function of \snls. From this fit, we determine that a fidelity of 90\% is reached at \snls~$=5.7$. This implies that for every ten sources with \snls~$= 5.7$, roughly one source will be due to ``noise'' fluctuations within the data. For \snls~$ \geq 6.0$, none of the data sets show any negative emission at this level, and the fit yields a $>98$\% probability that an emission line with this \snls\ is real.

In Appendix \ref{sec:autoline_det} we provide a detailed description of the line detections based on the LineSeeker searches. To summarize, a total of 10 emission lines were detected in the 15 sightlines with \snls~$\geq 5.7$. Four of these are identified as far-infrared emission lines associated with the quasar, while five of the emission lines are identified as \CII\ emission from galaxies associated with our target DLAs. Only one emission line of the ten remains unidentified. This detection rate is similar to that of previous studies of serendipitous line emitters in large ALMA programs \citep[e.g.,][]{Venemans2020,Decarli2020,Loiacono2021}.

\subsection{Visual search for line emission at the DLA redshift}
\label{sec:manline}

Previous works on searches for DLA galaxies \citep[e.g.,][]{Neeleman2017,Neeleman2019,Kanekar2020} have relied on visually inspecting the spectral cubes covering the DLA redshift to look for emission features. By restricting the spectral search range (by about a factor of 8, i.e. within $\pm 500$~\kms of the DLA redshift) compared to the full velocity range of the cube, these searches are able to provide more stringent S/N thresholds. To quantify this, we can determine the likelihood that a spurious high S/N noise feature falls within the above restricted search volume around the DLA. Using the automated search algorithm that ran on the full search volume, we sum the number of false negative sources and multiply this by the fractional difference in the search volume. From this, we expect that we find about $1 \pm 1$ sources with \snls~$> 5.4$ and about $8\pm3$ sources with \snls~$> 5.0$ in the combined 16 DLA fields. To minimize the number of false positives, we therefore only consider sources that have S/N~$> 5.0$ in our visual inspection. In passing, we emphasize that the S/N calculation in LineSeeker differs slightly from the manual S/N calculation due to differences in channel binning and the handling of correlated noise. In general, the manual S/N ratio is slightly higher. Strictly selecting sources with a manual S/N $> 5$ would therefore yield slightly more false positives than the above estimates, but removing very narrow (widths of $1-2$ channels) features largely offsets this increase in false positives.

For the visual inspection, we use the two independently reduced data sets described in Section \ref{sec:sample}, and search for features with S/N~$> 5$ in both sets. The full results are tabulated in Appendix~\ref{sec:manline_det}. To summarize, we find a total of 17 line candidates that have S/N~$> 5$ in either of the data sets. Of these 17 line candidates, seven are seen in both data sets. The remaining 10 line candidates (four and six in the independent sets, respectively) have S/N~$> 5$ in one, but not in the other set. The seven line candidates identified in both data sets consist of the five emission lines also found in the automated search algorithm (Sec.~\ref{sec:autoline}), and two additional lower-S/N line candidates. These two line candidates were previously reported in \citet{Neeleman2019} and \citet{Kaur2021} for the line emitters in PSS1443$+$27 and J1054$+$1633, respectively. Their relatively large line widths and proximity to another line emitter at the same redshift suggest that they are likely real. On the other hand, the line emission associated with the $z=4.3446$ DLA toward J1101$+$0531, reported by \citet{Neeleman2019}, may be spurious. For this paper, we will classify the seven line candidates recovered in both data sets as detections. The remaining 10 features with S/N~$> 5$ are classified as line candidates, and will not be used for any of the later analysis; these systems should be followed up for confirmation. Overall, this implies that we have detected seven line emitters in the fields of five DLAs at $z \approx 4.11-4.5$. 

\subsection{Search for continuum emission}
\label{sec:contsearch}

We also searched our continuum images for sources detected in the rest-frame far-infrared continuum. Here, we define a detection as a continuum source with flux density such that the S/N~$> 4.4$. This S/N detection limit is lower than that used for the line features in the spectral cubes, because there are fewer independent resolution elements in the continuum image compared to the spectral cube \citep[see, e.g.,][]{Gonzalez-Lopez2019}. For these searches, we assume that the continuum sources are unresolved, which is a good approximation at the observing frequencies and angular resolutions of our survey. For Gaussian noise, we expect a total of $< 0.1$ continuum detections with S/N~$> 4.4$ in the 15 continuum fields of this survey due to random noise fluctuations \citep[see, e.g.,][]{Condon1998, Dunlop2017}. Our assumed noise threshold is thus relatively conservative, and yields a high probability that any continuum detections are real. Details of the continuum detections are listed in Appendix~\ref{sec:contdet}. Briefly, all 14 target quasars are detected in the continuum, along with 14 additional continuum sources. Two of these continuum sources lie at the same location as the line emitters, and we assign this continuum emission to these line emitters (see Section~\ref{sec:firfromdla}). The 12 serendipitous continuum detections are consistent with expectations from blind surveys such as ASPECS \citep{Gonzalez-Lopez2020}.

\section{Analysis}
\label{sec:analysis}

In this section, we analyze the results from our search for redshifted \CII\ line emission (and associated continuum emission) from the DLA redshifts. The analysis of the other emission lines, which are primarily from the quasars within the targeted fields, is presented in Appendix~\ref{sec:qsoanalysis}.

\subsection{The galaxies associated with DLAs}
\label{sec:associateddlagal}

\begin{figure*}[!tbh]
\includegraphics[width=\textwidth]{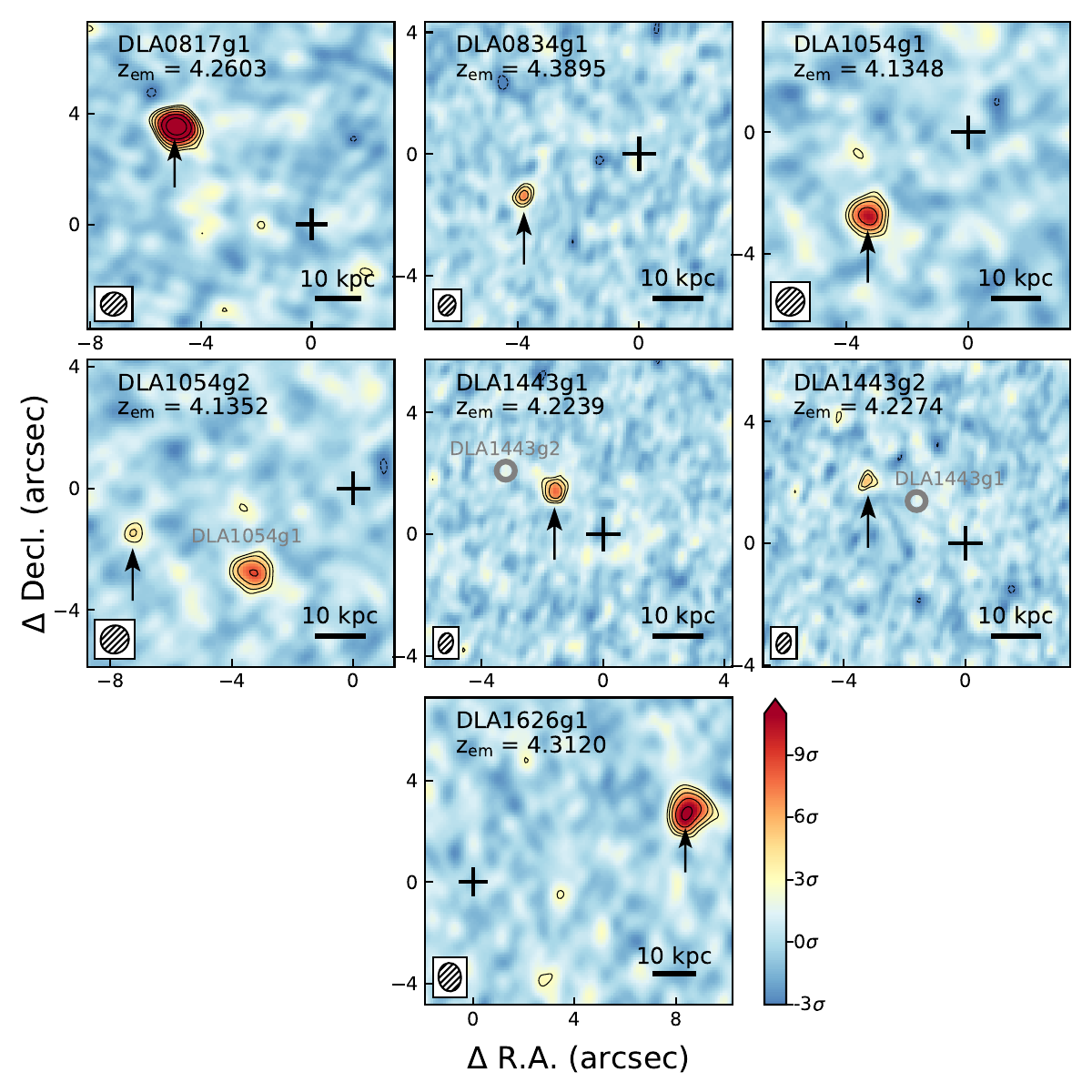}
\caption{Velocity-integrated \CII\ line emission for the seven associated DLA galaxies. The velocity range chosen maximizes the S/N of the \CII\ emission line for that galaxy. For DLA1054g2, the associated DLA galaxy DLA1054g1 is also seen due to their similar redshift. For DLA1443g2, the location of DLA1443g1 is marked by a gray circle, and vice versa. The plus sign marks the position of the DLA (and quasar) and the arrow marks the associated DLA galaxy. The beam for each of the observations is shown in the bottom left inset. The contours in each panel start at 3$\sigma$ and increase by powers of $\sqrt{2}$. The impact parameters between the DLAs and the associated DLA galaxies range between 14~kpc and 59~kpc.
\label{fig:dlagal}}
\end{figure*}

\begin{figure*}[!tbh]
\includegraphics[width=\textwidth]{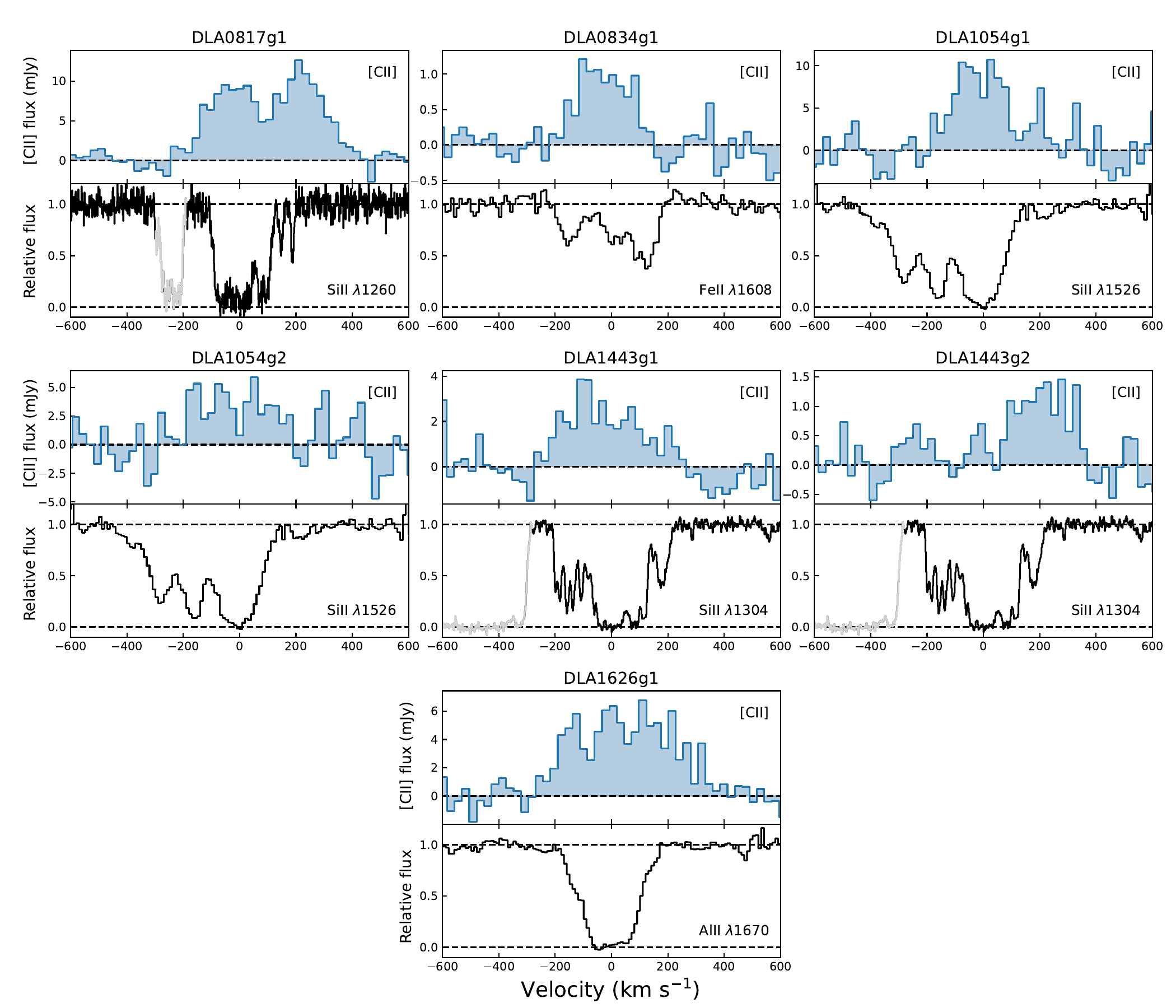}
\caption{Comparison between the \CII\ spectra of the associated DLA galaxy and the metal absorption for the DLA. The \CII\ spectra are either from a 1$''$ radius region for fields with resolved \CII\ emission (Table~\ref{tab:firprop}) or from the peak pixel in the velocity-integrated \CII\ maps for unresolved \CII\ emission (Fig.~\ref{fig:dlagal}). The metal line used for the comparison is noted in the bottom panel for each of the associated DLAs. Two of the absorption spectra (DLA0817g1 and DLA1443g1/DLA1443g2) are from the Keck/HIRES instrument, which has a velocity resolution of $\approx 6$~\kms. The remaining spectra are from Keck/ESI, which has a velocity resolution of 44~\kms. The velocity is relative to the DLA redshift (see Table~\ref{tab:dlaprop}). Absolute velocity differences between the redshift of the associated DLA galaxy and that of the DLA range between 10~\kms\ and 190~\kms. 
\label{fig:absems}}
\end{figure*}

When searching for emission from galaxies at the DLA redshift, it is important to first define the nature of the association. By definition, all galaxies identified at or around the DLA redshift by searches 
targeting the DLA redshift are \HI-selected galaxies or DLA galaxies. However, this does not necessarily mean that the identified galaxies have a direct connection with the absorbing gas, i.e. with the DLA. If a DLA arises in either the ISM or the CGM of a galaxy, i.e. if the impact parameter is smaller than the virial radius of the galaxy, the galaxy is expected to influence the properties of the DLA (e.g. the DLA's \HI\ column density, metallicity, element abundances, etc). This assumption is corroborated by simulations, which show that gas within the virial radius is influenced by the galaxy via outflows, and also show that such gas plays an integral part in the subsequent evolution of the galaxy  \citep[e.g.,][]{Muratov2015,Stern2021}. We refer to such galaxies, with DLA impact parameters within the virial radius,  as ``associated DLA galaxies''.

For DLAs arising in the ISM, the velocity offset between the emission and absorption redshifts is $\lesssim$200~\kms\ and the spatial separation $\lesssim$50~kpc, even for massive galaxies \citep[based on the local \HI\ mass-size relation and typical galaxy rotation velocities; e.g. ][]{Wang2016}. Conversely, if the DLA arises in the CGM of a massive galaxy, the spatial separation between the DLA sightline and the galaxy is likely to be $\lesssim$100~kpc and the velocity offset $\lesssim$500~\kms. The above estimates are redshift dependent and assume a redshift of $z \sim 4$; they are broadly consistent with expectations from clustering analyses and numerical simulations \citep[e.g.][]{Bird2014,Perez-Rafols2018a}. Further, the virial radius of a halo of mass $10^{11} \, M_\odot$ at $z \approx 4$ is $\approx$ 100~kpc \citep[e.g.][]{Munoz-Cuartas2011}. We will hence consider galaxies identified within $\approx \pm 500$~\kms\ from the DLA redshift and with impact parameters $\lesssim 100$~kpc from the DLA sightline as being directly associated with the $z \sim 4$ DLAs. 

For the present observations, the ALMA primary beam at the target frequencies has a half-power radius of $\approx 8\farcs3$, i.e. a physical size of $\approx$ 60~kpc at $z \approx 4$. As such, in terms of spatial separation, our observations are only sensitive to emission from galaxies that are directly associated with the target DLAs. This is unlike, for example, the \lya\ searches of \citet{Lofthouse2023}, which have a much larger field of view.

To summarize, we consider any emission line that falls within the ALMA primary beam and whose measured frequency is within $\approx \pm 500$~\kms\ of the redshifted \CII\ line frequency, to be a \CII\ emission line from a galaxy associated with the DLA. We note, based on results from previous ``blind'' surveys \citep[e.g.][]{Decarli2020}, that the likelihood of a lower-redshift emission line being incorrectly identified as a \CII\ emission line at the DLA redshift is negligible. Similarly, we find only one unidentified line in the spectral windows not associated with the redshifted \CII\ emission in our full sample, further emphasizing that serendipitous lines are rare in ALMA observations. 

With these criteria in place, we can classify all seven line emitters detected in our survey as associated DLA galaxies. Two of the DLAs, PSS1443+27 and J1054+1633, have two associated DLA galaxies close to the DLA redshift. Figure \ref{fig:dlagal} shows images of the velocity-integrated \CII\ emission from the seven associated DLA galaxies, while Figure~\ref{fig:absems} shows the \CII\ spectra of the 7 galaxies along with a spectrum of a representative metal absorption line in each DLA.

\subsection{[CII] and FIR continuum emission from the DLA fields}
\label{sec:firfromdla}

\begin{deluxetable*}{lllllllllcc}
\tablecaption{\CII\ and $160\mu$m continuum properties of the DLA galaxies
\label{tab:firprop}}
\tablehead{
\colhead{ID} &
\colhead{z$_{\rm em}$} &
\colhead{FWHM} &
\colhead{$\int S_\nu dv$} &
\colhead{$L_{\rm [CII]}$} &
\colhead{$S_{\rm 160\mu m}$} &
\colhead{$L_{\rm 160\mu m}$} &
\colhead{$L_{\rm TIR}$} &
\colhead{SFR$_{\rm 160\mu m}$} &
\colhead{$b$} &
\colhead{$\Delta v_{\rm D-G}$}\\
\colhead{} &
\colhead{} &
\colhead{(\kms)} &
\colhead{(Jy \kms)} &
\colhead{($10^8 M_\odot$)} &
\colhead{(mJy)} &
\colhead{($10^{10} M_\odot$)} &
\colhead{($10^{11} M_\odot$)} &
\colhead{($M_\odot$ yr$^{-1}$)} &
\colhead{(kpc)} &
\colhead{(\kms)}
}
\startdata
DLA0817g1$^{\rm a}$&4.2603&431 $\pm$ 23&4.18 $\pm$ 0.19&23.5 $\pm$ 1.1&1.18 $\pm$ 0.10&19.8 $\pm$  1.6&10.6 $\pm$  5.4&108 $\pm$ 44&41&-108\\
DLA0834g1&4.3895&209 $\pm$ 31&0.24 $\pm$ 0.03&1.4 $\pm$ 0.2&$<$0.06&$<$ 1.1&$<$ 0.7&$<$ 7&27& 28\\
DLA1054g1$^{\rm a}$&4.1348&233 $\pm$ 32&2.36 $\pm$ 0.25&12.6 $\pm$ 1.4&0.23 $\pm$ 0.07& 3.8 $\pm$  1.1& 2.0 $\pm$  1.2&21 $\pm$ 10&29&-10\\
DLA1054g2&4.1352&327 $\pm$ 73&1.41 $\pm$ 0.25&7.5 $\pm$ 1.3&$<$0.31&$<$ 4.9&$<$ 3.2&$<$30&51&-34\\
DLA1443g1$^{\rm a}$&4.2239&288 $\pm$ 54&0.87 $\pm$ 0.13&4.8 $\pm$ 0.7&0.19 $\pm$ 0.09& 3.1 $\pm$  1.5& 1.7 $\pm$  1.1&17 $\pm$ 10&14& 11\\
DLA1443g2&4.2274&228 $\pm$ 43&0.31 $\pm$ 0.05&1.7 $\pm$ 0.3&$<$0.10&$<$ 1.7&$<$ 1.1&$<$10&26&-190\\
DLA1626g1$^{\rm a}$&4.3120&429 $\pm$ 42&2.51 $\pm$ 0.21&14.4 $\pm$ 1.2&0.30 $\pm$ 0.06& 5.1 $\pm$  1.0& 2.8 $\pm$  1.5&28 $\pm$ 13&59&-57\\
\tableline
J0307-4945F&4.4679&200&$<$0.05&$<$0.3&$<$0.05&$<$ 0.9&$<$ 0.6&$<$ 6&\nodata&\nodata\\
J0817+1351F&4.2584&200&$<$0.10&$<$0.6&$<$0.11&$<$ 1.9&$<$ 1.2&$<$11&\nodata&\nodata\\
J0824+1302F&4.4720&200&$<$0.06&$<$0.4&$<$0.07&$<$ 1.4&$<$ 0.9&$<$ 8&\nodata&\nodata\\
J0834+2140F&4.3900&200&$<$0.05&$<$0.3&$<$0.06&$<$ 1.0&$<$ 0.6&$<$ 6&\nodata&\nodata\\
J0834+2140F2&4.4610&200&$<$0.07&$<$0.4&$<$0.07&$<$ 1.2&$<$ 0.8&$<$ 8&\nodata&\nodata\\
BR0951-04F&4.2029&200&$<$0.08&$<$0.4&$<$0.07&$<$ 1.1&$<$ 0.7&$<$ 7&\nodata&\nodata\\
J1054+1633F&4.1346&200&$<$0.23&$<$1.3&$<$0.17&$<$ 2.8&$<$ 1.8&$<$17&\nodata&\nodata\\
J1100+1122F&4.3947&200&$<$0.06&$<$0.3&$<$0.04&$<$ 0.8&$<$ 0.5&$<$ 5&\nodata&\nodata\\
J1101+0531F&4.3446&200&$<$0.09&$<$0.5&$<$0.09&$<$ 1.5&$<$ 1.0&$<$ 9&\nodata&\nodata\\
J1132+1209F&4.3802&200&$<$0.08&$<$0.4&$<$0.07&$<$ 1.2&$<$ 0.8&$<$ 7&\nodata&\nodata\\
BR1202-07F&4.3829&200&$<$0.09&$<$0.5&$<$0.10&$<$ 1.7&$<$ 1.1&$<$10&\nodata&\nodata\\
PSS1443+27F&4.2241&200&$<$0.08&$<$0.4&$<$0.09&$<$ 1.4&$<$ 0.9&$<$ 9&\nodata&\nodata\\
J1607+1604F&4.4741&200&$<$0.06&$<$0.4&$<$0.06&$<$ 1.1&$<$ 0.7&$<$ 7&\nodata&\nodata\\
J1626+2751F&4.3110&200&$<$0.08&$<$0.5&$<$0.08&$<$ 1.3&$<$ 0.9&$<$ 8&\nodata&\nodata\\
J1626+2751F2&4.4975&200&$<$0.06&$<$0.4&$<$0.08&$<$ 1.4&$<$ 0.9&$<$ 9&\nodata&\nodata\\
PSS2241+1352F&4.2833&200&$<$0.09&$<$0.5&$<$0.09&$<$ 1.5&$<$ 1.0&$<$ 9&\nodata&\nodata
\enddata
\tablecomments{Table is subdivided into \CII\ detections (above the horizontal line) and limits on any galaxies within the ALMA seach area for all 16 DLA fields (below the horizontal line). For the latter, the 3$\sigma$ upper limits on $\int S_\nu dv$ and $L_{\rm [CII]}$ assume a line FWHM of 200 \kms\ and a redshift equal to that of the DLA. For $L_{\rm TIR}$, the uncertainties include an additional 0.5 dex to account for the uncertianties in the black body parameters. The SFR includes a 0.2 dex uncertainty to account for the scatter in the relation between SFR and $L_{\rm 160\mu m}$. $\Delta v_{\rm D-G}$ is the velocity difference between the redshift of the DLA and the associated DLA galaxy with typical uncertainties of $\approx$15 \kms. Finally, $b$ denotes the impact parameter between the DLA sightline and the galaxy.\\$^{\rm a}$ Galaxies were spatially resolved in the ALMA observations. For these galaxies, the continuum and \CII\ flux densities are derived from a circular region with a radius of 1$''$ centered on the emission.}
\end{deluxetable*}

For the seven detections of redshifted \CII\ emission, Table~\ref{tab:firprop} lists the velocity-integrated \CII\ line flux density, $\int S_\nu d\nu$, the \CII\ line luminosity, $L_{\rm [CII]}$, the $160\mu$m continuum flux density, $S_{\rm 160 \mu m}$, and the monochromatic $160\mu$m continuum luminosity, $L_{\rm 160 \mu m}$. For two of the \CII\ emitters, the continuum was found in the continuum search described in Section \ref{sec:contsearch}. For the remaining \CII\ emitters, one has its continuum detected at $> 3\sigma$ significance, and, another at $\approx 2\sigma$ significance. Because of the excellent spatial agreement, we consider these detections. The remaining three \CII\ emitters are not detected in the continuum. In Table~\ref{tab:firprop}, we further list the 3$\sigma$ limits for all of the 16 DLA fields, where we have assumed a \CII\ line FWHM of 200~\kms\ for estimating $\int S_\nu d\nu$ and $L_{\rm [CII]}$.

The measurements and upper limits of $S_{\rm 160 \mu m}$ have been used to estimate the total FIR luminosity of the associated DLA galaxies, $L_{\rm TIR}$, which is defined as the luminosity emitted between 8 and 1000~$\mu$m. For this, we assume that each galaxy has a modified black-body spectrum with a dust temperature of 35~K, and with $\beta=1.6$ \citep[e.g.,][]{Carilli2013,Neeleman2017,Faisst2020}. These choices are based on the typical conditions of main-sequence galaxies at these redshifts. However, the values of these quantities have been shown vary significantly in the few studies that have been able to constrain them at $z \approx 4$ \citep[e.g.,][]{Faisst2020}. We therefore conservatively add a 0.5~dex systematic uncertainty to $L_{\rm TIR}$ to account for the uncertainty in dust temperature and $\beta$. The $160\mu$m continuum flux density can also be used to estimate the SFR of a galaxy, by assuming that $L_{\rm 160 \mu m}$ scales with SFR as seen in local galaxies \citep{Calzetti2010}. The inferred SFR$_{\rm 160\mu m}$ values or upper limits are listed in Column~9 of Table~\ref{tab:firprop}. For the four detections of $160\mu$m continuum emission, the SFR value lies in the range $17-108$~\msunyr, while, for the rest of the galaxies, our $3\sigma$ upper limits are typically $< 10$~\msunyr. We note that there is a large scatter in the scaling between $L_{\rm 160 \mu m}$ and SFR, which we take into account by adding  0.2~dex to the SFR estimate uncertainties listed in Table \ref{tab:firprop}.

For the seven identified associated DLA galaxies, the last two columns of Table \ref{tab:firprop} list the impact parameter between the DLA and the galaxy, $b$, and the velocity difference between the flux-weighted centroid of the \CII\ emission in the DLA galaxy, and the optical-depth-weighted centroid of the low-ion absorption lines in the DLA,  $\Delta v_{\rm D - G}$.  A negative velocity corresponds to a DLA that has a lower redshift than that of the DLA galaxy. The impact parameters lie in the range $14-59$~kpc, while the maximum absolute value of the velocity difference is $190$~\kms. 

\section{Discussion}
\label{sec:discussion}

\subsection{[CII] and FIR properties of associated DLA galaxies}

\begin{figure*}[!tbh]
\includegraphics[width=0.5\textwidth]{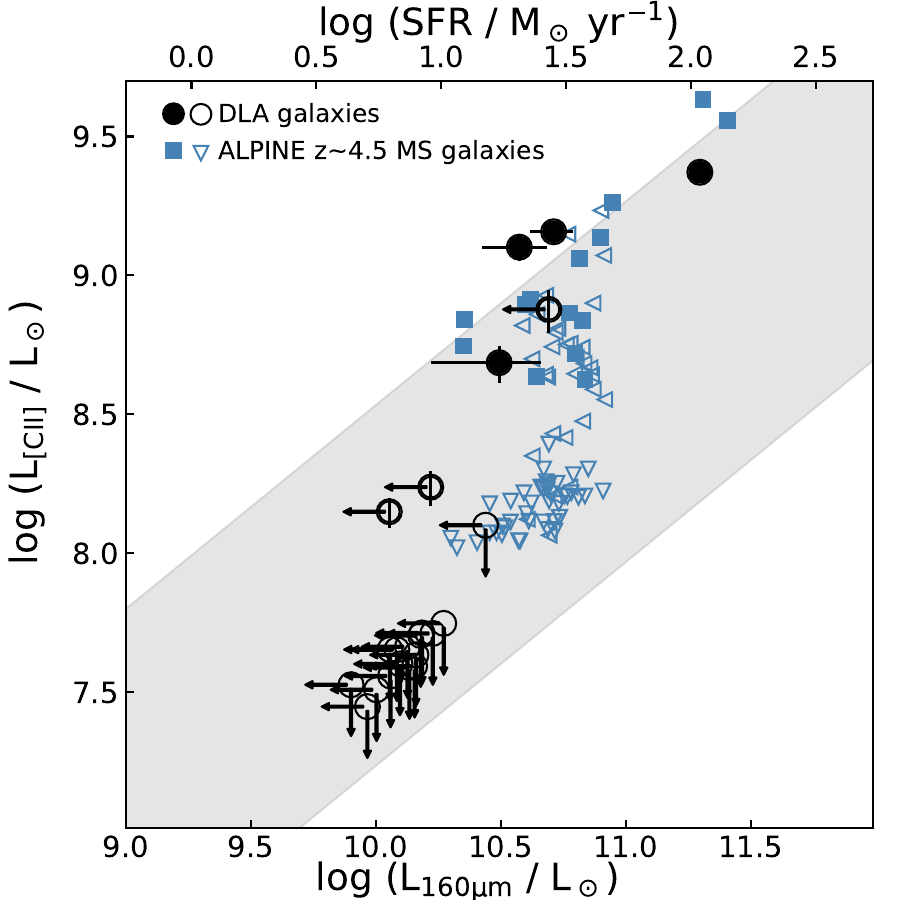}
\includegraphics[width=0.5\textwidth]{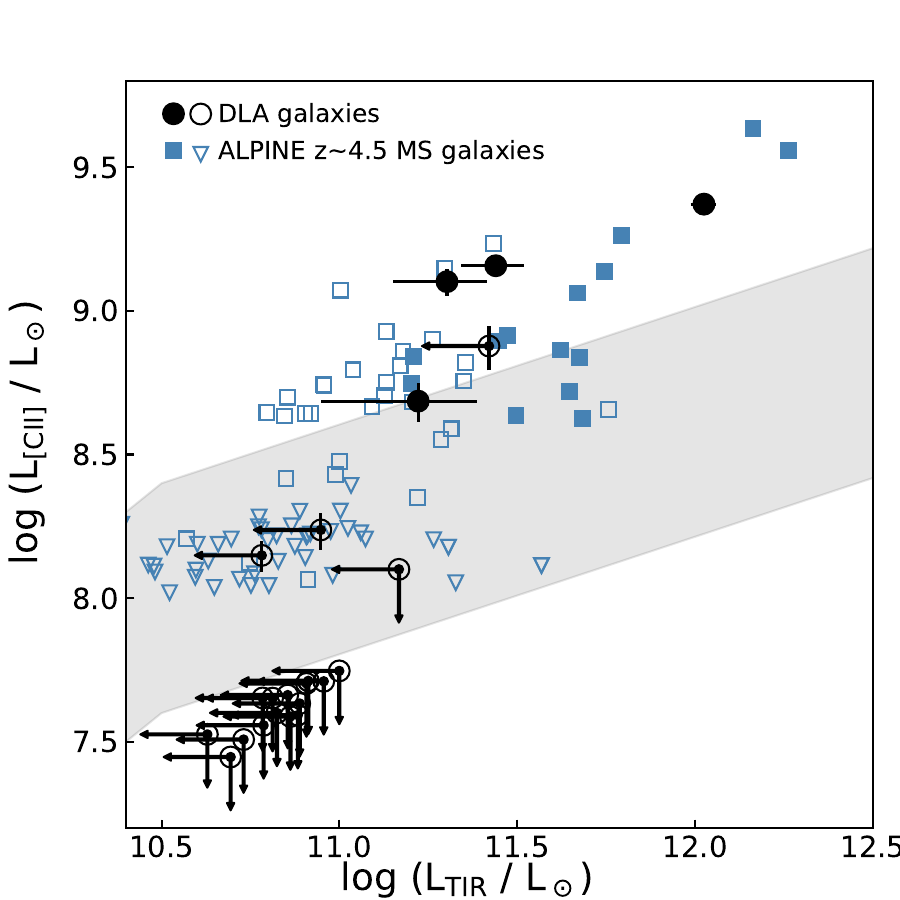}
\caption{Left: \CII\ luminosity as a function of the far-infrared dust continuum luminosity measured at a rest-frame wavelength of 160$\mu$m. The top axis shows the inferred SFR, assuming the scaling relation of \citet{Calzetti2010}. Right: ratio of the \CII\ luminosity and total far-infrared dust continuum luminosity as a function of the total far-infrared luminosity, i.e., the so-called \CII-deficit. The gray band shows the parameter space occupied by local galaxies. The solid symbols indicate detections in both \CII\ and FIR continuum, whereas open symbols indicate limits, where the triangles indicate the direction of the respective limit. The associated DLA galaxies have FIR properties that are consistent with those of other observed high-redshift galaxies, while the limits (3$\sigma$) of our observations are about 4$\times$ fainter than typical for high-redshift galaxy observations.
\label{fig:ciifir}}
\end{figure*}

One of the primary aims of our survey is to characterize the galaxies associated with DLAs, and to place them in the context of other galaxy populations at similar redshifts. For the latter, we use galaxies of the ALMA Large Program to INvestigate \CII\ at Early times (ALPINE) survey as our comparison sample \citep{Lefevre2020,Bethermin2020}. The ALPINE survey observed 118 main-sequence galaxies at $z \approx 4.4 - 5.9$ with ALMA in \CII\ emission, with 67 galaxies at $z \approx 4.4-4.65$ and 51 at $z \approx 5.1-5.9$ \citep{Faisst2020,Bethermin2020}. The 67 ALPINE galaxies at $z \approx 4.5$ lie at very similar redshifts to our target DLAs, and form a uniformly-selected sample; we hence use this subsample as our comparison sample. We note that the ALPINE survey is significantly shallower than our \CII\ survey: the ALPINE on-source times are 15m--25m per source \citep{Bethermin2020}, while our on-source times are $\approx$~0.75h--3.1h per source (see Table~\ref{tab:obs}).

In the left panel of Figure \ref{fig:ciifir}, we plot $L_{\rm [CII]}$ against $L_{\rm 160\mu m}$ for our associated DLA galaxies and the ALPINE comparison sample. The well-known correlation between these two observables has been explained as arising due to a correlation between $L_{\rm [CII]}$ and SFR \citep{Boselli2002,DeLooze2014,Schaerer2020}, because $L_{\rm 160\mu m}$ is known to correlate with the SFR \citep{Calzetti2010}. We find that the measured \CII\ and $160\mu$m luminosities of our associated DLA galaxies at $z \approx 4.1 - 4.5$ are in excellent agreement with the corresponding measurements for the main-sequence galaxies of the ALPINE survey at $z \approx 4.5$ (although our \CII\ detections extend to lower luminosities, log$\rm [L_{[CII]}/L_\odot] \approx 8$).

In the right panel of Figure \ref{fig:ciifir}, we plot $L_{\rm [CII]}$ against $L_{\rm TIR}$ for the DLA galaxies and the ALPINE galaxies at $z \approx 4.5$. For the ALPINE galaxies, we use the IRX-$\beta$ relation between the IR excess and the UV spectral slope to estimate $L_{\rm TIR}$ for those galaxies not directly detected in the FIR continuum. \citep{Schaerer2020,Fudamoto2020}. We also plot the range of measurements for galaxies in the local Universe \citep[e.g.][]{Malhotra1997,Malhotra2001,Luhmann1998}. In the case of the ALPINE main-sequence sample, it is clear that most of the galaxies that show \CII\ detections lie above the region occupied by $z \approx 0$ galaxies with similar $L_{\rm TIR}$ values. Similarly, five of the seven DLA galaxies also lie above this region and are fully consistent with the ALPINE sample, while the other two systems are not detected in the FIR continuum. The higher $L_{\rm [CII]}$ for both the DLA galaxies and the ALPINE galaxies suggest that these high redshift galaxies may have colder dust temperatures than $z \approx 0$ galaxies with similar $L_{\rm TIR}$ values.  Overall, the \CII\ and FIR continuum properties of our DLA galaxies are similar to those of main-sequence galaxies at $z \approx 4.5$.

Finally, the relatively high sensitivity of our survey implies that those DLA galaxies that are not detected in dust continuum emission, as well as the DLA fields without \CII\ detections, have upper limits on $L_{\rm [CII]}$ and $L_{\rm 160\mu m}$ that are roughly 0.5 dex lower than the constraints from large high-$z$ surveys like the ALPINE survey \citep[e.g.,][]{Schaerer2020}. These limits indicate that the dominant population of galaxies probed by DLAs has \CII\ luminosities $\lesssim 3 \times 10^7 \,L_\odot$, and SFR $\lesssim 10$ \msunyr. If DLA galaxies lie on the star-forming main sequence at $z \approx 4.3$ \citep[e.g.][]{Popesso2023}, our SFR constraints imply stellar masses $\lesssim 2 \times 10^{9} \, M_\odot$. These are consistent with expectations from simulations \citep[e.g.,][]{Pontzen2008,Bird2014,DiGioia2020} and inferences from observations \citep{Perez-Rafols2018a} about the typical mass of the galaxies associated with high-redshift DLAs. Overall, we find that the \CII\ and FIR properties of our DLA galaxies are consistent with a scenario in which most DLAs at $z \approx 4.3$ are associated with faint main-sequence galaxies.

\subsection{The DLA -- galaxy connection at $z \approx 4$}
\label{sec:dlagal}

\begin{figure*}[!tbh]
\includegraphics[width=0.5\textwidth]{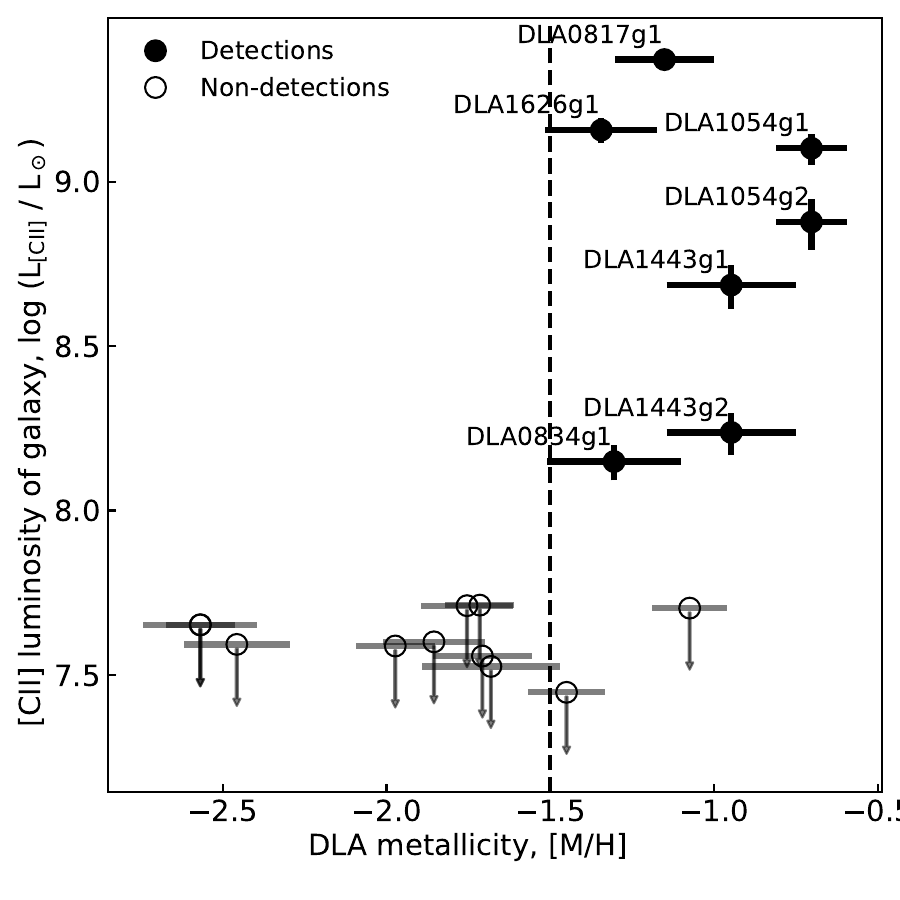}
\includegraphics[width=0.5\textwidth]{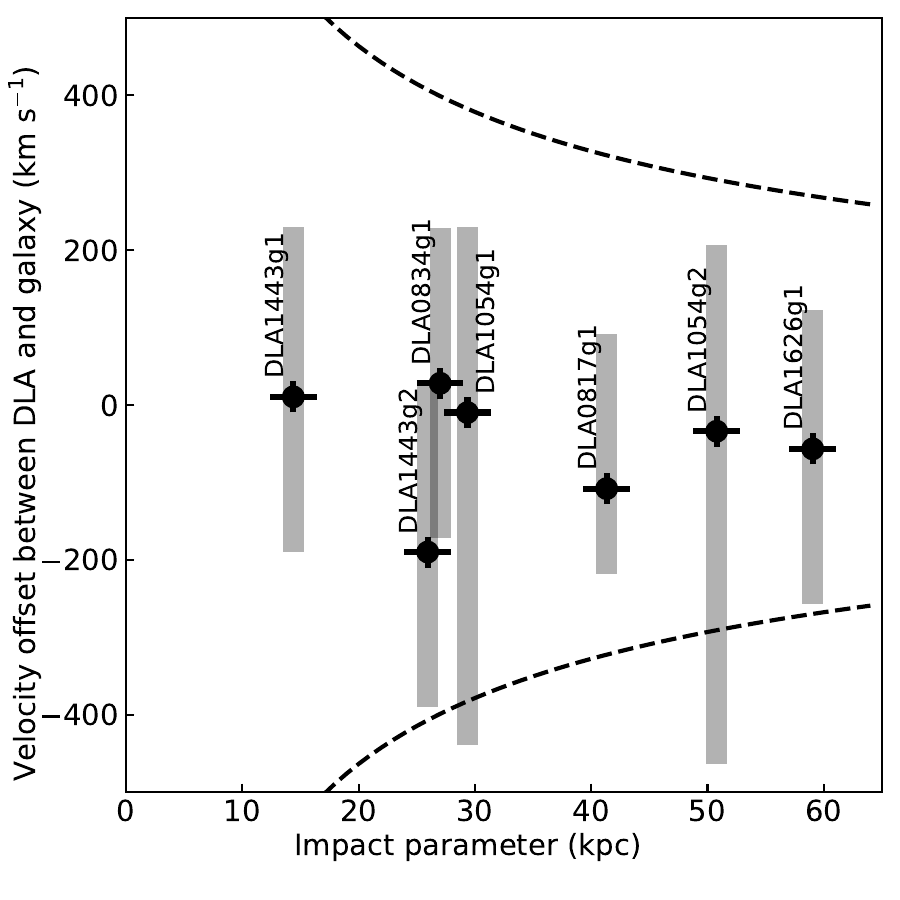}
\caption{Left: \CII\ luminosity of DLA galaxies as a function of the metallicity of the DLA. Non-detections are 3$\sigma$ upper limits. The vertical line corresponds to a metallicity of [M/H] = -1.5. In this paper, we will refer to DLAs with metallicities above this threshold as 'high-metallicity' DLAs, and those below as 'low-metallicity' DLAs. We stress that these are subjective terms and what is considered high- or low-metallicity will vary as a function of both redshift and scientific intent. We find that the fields of high-metallicity DLAs typically contain at least one \CII-emitting galaxy. Right: Velocity offset between the DLA and the associated galaxy as a function of impact parameter. The vertical gray bars correspond to the full range of absorption seen in the DLA. For reference, we also plot the escape velocity of a $10^{12}~M_\odot$ galaxy, where for simplicity we assume all of the mass is within the measured impact parameter \citep[see;][for a more detailed analysis]{Christensen2019}. The small difference in velocities between the DLA and DLA galaxy indicates that the bulk of the material probed by the DLA is gravitationally bound to the associated DLA galaxies.
\label{fig:dlagalcomp}}
\end{figure*}

Out of our sample of 16 DLAs at $z \approx 4.1-4.5$, five DLAs have at least one associated galaxy that was detected in \CII\ emission. The left panel of Figure~\ref{fig:dlagalcomp} plots the \CII\ line luminosity of the associated galaxy versus the DLA metallicity: it is clear that all \CII-emitting galaxies are found in the fields of DLAs with [M/H]~$>-1.5$ (indicated by the dashed vertical line in the figure). The \CII\ detection rate is $71^{+11}_{-20}$\% for DLAs with [M/H]~$> -1.5$, and $0^{+18}$\% for DLAs with [M/H]~$< -1.5$. Here the uncertainties are 68\% confidence intervals derived from small number statistics by taking into account the finite sampling of the data set. To be specific, we calculate the probability distribution function that a given rate yields the observed detection rate, and report the 68\% confidence limit on this distribution. We will henceforth refer to the DLAs of our sample with [M/H]~$> -1.5$ as high-metallicity DLAs, and those with [M/H]~$< -1.5$ as low-metallicity DLAs. We stress that these are subjective terms, and we apply these only to the DLAs of our sample. 

The left panel of Figure~\ref{fig:dlagalcomp} yields two immediate conclusions. First, observing the fields of high-metallicity DLAs at $z \approx 4$ with ALMA is an efficient way to uncover a population of \CII-emitting galaxies that are not selected by their emission properties. This is in contrast to deep ``blind'' surveys such as ASPECS, which have shown that $z>4$ \CII\ emitters are difficult to detect in the small cosmological volumes probed by these surveys \citep{Walter2016,Decarli2020}. As a result, our current knowledge of \CII-emitting galaxies at high redshifts is driven by preselecting galaxies based on their emission properties. This introduces biases that are typically difficult to quantify.  The absorption selection employed in this paper highlights a complementary approach to select high-redshift galaxies without preselecting based on their emission properties. 

Second, high-metallicity DLAs are more likely than low-metallicity DLAs to have an associated luminous \CII-emitting galaxy. This might not appear surprising, given the well-known correlation between galaxy metallicities and stellar masses at all redshifts \citep[e.g.,][]{Tremonti2004,Erb2006,Mannucci2009,Sanders2018,Maiolino2019,Henry2021,Revalski2024} and the fact that massive galaxies are expected to have higher SFRs and thus higher \CII\ luminosities \citep{Popesso2023,Schaerer2020}. However, the impact parameters of the \CII-emitting galaxies to the DLA sightlines are $\approx 14-60$~kpc, with a median of $\approx 29$~kpc. This implies that the DLA sightline is probing gas outside the ISM of these galaxies, and that despite the large impact parameter, the DLA metallicity is a good tracer of the \CII\ luminosity of the nearby galaxy. If we assume that our galaxies lie on the galaxy main sequence \citep{Popesso2023} and naively assume that the DLA metallicity equals the metallicity of the DLA galaxy, then we find that the galaxies are consistent with (albeit slightly below) the mass--metallicity relation at this redshift \citep{Curti2024}.

The above connection between DLA metallicity and \CII\ luminosity can either be direct, i.e., the metals in the DLA originate from the galaxy itself, or indirect, i.e., the metals originate from a satellite galaxy of the \CII-emitting galaxy. We first consider the second possibility, that the DLA absorption arises in a satellite of the \CII-emitting galaxy. 

If the metals seen in the DLA arise from a satellite galaxy, we would expect to see a velocity offset between the low-ionization metal absorption and \CII\ emission lines, due to the motion of the satellite galaxy relative to the \CII\ emitter. Figure~\ref{fig:absems} shows a comparison between the \CII\ emission and metal-line absorption profiles for each DLA, while the last column of Table~\ref{tab:firprop} lists the velocity difference $\Delta v_{\rm D - G}$ between the DLA and the DLA galaxy. The right panel of Figure~\ref{fig:dlagalcomp} shows the velocity difference plotted as a function of DLA impact parameter where the gray bars indicate the velocity range of all of the gas components seen in absorption. This figure shows that the velocity difference between mean emission and mean absorption is relatively small, with no dependence on the impact parameter. To provide a quantitative comparison on the magnitude of the mean velocity offset, we compare this to the typical mean velocity offset of globular clusters in our Milky Way. For the five DLAs with identified \CII\ emitters\footnote{For the two DLAs with two identified \CII\ emitters in the field, we use the closer \CII\ emitter in this comparison.}, we computed the average radial velocity offset between absorption and emission lines, defined as,
\begin{equation}
\Delta v_{\rm rad} = \frac{\sum^N_i |{\Delta v_{{\rm D - G}, i}}|}{N} \, .
\end{equation}
We obtain an average radial velocity offset of $60^{+30}_{-10}$~\kms, far smaller than the typical offsets expected for satellite galaxies around a massive galaxy. For example, the mean radial velocity offset for the satellite galaxies (and globular clusters) of the Milky Way is $114^{+11}_{-1}$~\kms \citep{Hartwick1978}. This small velocity offset is remarkable, because it implies that the mean gas component seen in absorption is tied better to these high-redshift galaxies than the globular clusters are tied to our Milky Way, even though the gas extends out to $\approx$60~kpc. One possible explanation is that the absorption sightline probes a large number of gas clumps within the CGM, and therefore already is averaged over many individual clumps. Regardless, the relatively small velocity offsets between absorption and emission lines make it very unlikely that the DLA absorption arises in satellites of the \CII-emitting galaxies.

Further, in the satellite galaxy scenario, one must account for the fact that the putative satellite galaxy remains undetected in \CII\ emission even in our relatively deep observations. We estimate the \CII\ luminosity of the putative satellite galaxy by conservatively assuming that its metallicity is equal to the DLA metallicity. Further, we assume that the satellite galaxy follows the mass--metallicity relation at $z \approx 4$ \citep{Mannucci2009,Maiolino2019}, and lies on the star-forming main sequence \citep{Popesso2023}. Given these assumptions, we find that a galaxy with a metallicity $\gtrsim -1.3$ would have an SFR $\gtrsim 8$~\msunyr. If we further assume that the galaxy follows the \CII--SFR scaling relation \citep[see Fig. \ref{fig:ciifir} and][]{Schaerer2020}, the expected \CII\ luminosity of the galaxy would be $L_{\rm [CII]} \gtrsim 5 \times 10^7 \, L_\odot$. Such galaxies would have been detected at $\gtrsim 4\sigma$ significance close to the DLA sightline in the majority of our observations.

We therefore assert that the metals seen in the DLA absorption do not arise from a satellite galaxy, but must arise from the \CII-emitting galaxy, either directly in a metal-rich outflow, or, more likely, after the metals have been mixed with the surrounding gas.

The conclusion that high-metallicity DLAs at $z \approx 4$  are tracing metals that arise from the associated \CII-emitting galaxies has two important implications. First, it implies that metals are ejected into the CGM out to large distances from luminous galaxies, $\approx 50-60$~kpc in the case of DLA1054g1 and DLA1626g1. This is consistent with some cosmological simulations that expect that metals are more easily ejected and mixed in high-$z$ galaxies compared to low-$z$ systems  \citep[e.g.][]{Muratov2015,Nelson2019}. The second implication is that high column densities of neutral gas are present in the CGM of high-redshift galaxies. All of the high-metallicity DLAs that were detected in \CII\ emission have \HI\ column densities of $N_{\rm HI} \approx 10^{21}$~cm$^{-2}$. In local galaxies, such high \HI\ column densities arise almost exclusively in the inner disks of galaxies, even in the case of interacting systems \citep[see e.g.,][]{DeBlok2018}. Our observations thus indicate that at $z \approx 4$, a significant amount of hydrogen in the CGM remains neutral, extending out to a significant fraction of the virial radius. This too has been seen in numerical simulations of high-$z$ galaxies \citep{Danovich2015,Stern2021}. In particular, \citet{Stern2021} suggest that gas can cool efficiently at high redshifts in lower mass halos, and that it remains neutral because of the high volume densities in these systems. 

It is important to note that our observations do not necessarily require that the \HI\ of the DLA itself originates in the galaxy seen in \CII\ emission, just that the bulk of the metals stem from this galaxy. The \HI\ could arise directly in the CGM, in outflowing gas from the galaxy, or even in dense clumps of inflowing gas, although our earlier arguments exclude the possibility of the \HI\ having large relative motions compared to the galaxy. Nevertheless, our observations do show that the metals must arise from the associated \CII-emitting DLA galaxy, and that the gas probed by the DLA is gravitationally bound to this galaxy.

We finish this section with a note on the detectability of DLA galaxies associated with low-metallicity DLAs. Our observations show that no \CII\ emission was found within the field of these DLAs, implying these DLA galaxies have SFR $< 8$~\msunyr. The depth of our ALMA observations implies that finding DLA galaxies associated with low-metallicity DLAs will be extremely challenging with current millimeter telescopes. 

\section{Summary and Conclusions}
\label{sec:summary}
We report results from an ALMA survey for redshifted \CII\ emission in the fields of DLAs at $z = 4.1-4.5$, aiming to identify and characterize the associated \HI-selected galaxies. The survey targeted 16 DLAs in the above redshift range, without pre-selection for either \HI\ column density or DLA metallicity, and thus targeted a representative range of DLA metallicities at $z \approx 4.3$. Using a combination of automated and visual searches, we detected \CII\ emission from seven galaxies in the fields of five DLAs; this constitutes our statistical sample of \CII\ detections. We detected four additional far-infrared emission lines from the background QSOs, and one unidentified emission line in the 15 spectral cubes, all at high ($> 5\sigma$) significance in two independent analyses. In addition, we identify 10 spectral features which have $> 5\sigma$ significance in one of the analyses, but $< 5\sigma$ in the other; these are classified as tentative detections, and not included in our statistical sample.

Based on this survey, we find that:
\begin{itemize}

\item There is a strong connection between the DLA metallicity and the presence of a \CII-emitting galaxy in the field, within $\approx 80$~kpc of the DLA sightline. For $z > 4$ DLAs with metallicity {M/H}~$> -1.5$, the \CII\ detection rate is $71^{+11}_{-20}$\%, whereas for DLAs below this metallicity threshold, the \CII\ detection rate drops to $0^{+18}$\%, where the uncertainties indicate 68\% confidence intervals. 

\item DLA galaxies at $z \approx 4$ have far-infrared properties that are consistent with those of $z \approx 4$ main-sequence galaxies, with typical stellar masses $\lesssim 10^{10} \, M_\odot$. The four DLA galaxies that are detected in dust continuum have inferred SFRs in the range $\approx 17-108$~\msunyr. The typical upper limits to the SFR for the remaining DLA galaxies are $\lesssim 10$~\msunyr. Together with the previous conclusion, this shows that ALMA observations of fields surrounding high-metallicity DLAs provide an efficient (in comparison with blind surveys) and complementary (in comparison to traditional surveys based on the emission properties of high-$z$ galaxies) approach to identifying normal, high-redshift, star-forming galaxies, without preselection based on the emission properties of the galaxy.

\item The velocity difference between the DLA and the DLA galaxy is relatively small, with absolute velocity offsets of $\approx 10 - 190$~\kms. Together with the large difference in detection rates between DLAs with high metallicity, [M/H]~$> -1.5$, and those with [M/H]~$< -1.5$, this indicates the metals along the DLA sightline must arise from the \CII-emitting galaxy.

\item The impact parameter between the DLA sightline and the \CII-emitting galaxy lies in the range $\approx 14 - 59$~kpc, with a median impact parameter of $29$~kpc. The relatively high metallicities, [M/H]~$> -1.5$, at these large impact parameters indicate that these regular, star-forming galaxies at $z \gtrsim 4$ have been very effective at enriching their CGMs out to a substantial fraction of the virial radius. Further, the large \HI\ column densities of the DLAs, N$_{\rm HI} \geq 10^{21}$~cm$^{-2}$, indicates that the CGMs of galaxies at $z \gtrsim 4$ contain far larger quantities of cold, neutral gas than the CGMs of galaxies at $z \approx 0$.
\end{itemize}

To summarize, our ALMA observations are consistent with a scenario where the \CII-emitting galaxy has typically enriched its surrounding CGM out to $\approx 30$~kpc, with enrichments out to distances of $50-60$~kpc in some cases. Within the CGM are clumps of dense \HI\ that have been enriched by the metals of the \CII-emitting galaxy. The high metallicity of these gas clumps is due to the metal enrichment from the associated galaxy, not from local star formation.  Unlike the metals, the origin of the \HI\ is not well-constrained; the \HI\ could stem from dense clumps of inflowing gas, from gas driven outwards by the \CII-emitting galaxy, from local gas cooling, etc. Multi-wavelength observations could help distinguish between these possibilities by providing a complementary view of the DLA fields at wavelengths unaffected by dust obscuration.

\begin{acknowledgments}
This paper makes use of the following ALMA data: ADS/JAO.ALMA\#2015.1.01564.S,\\ ADS/JAO.ALMA\#2016.1.00569.S,\\ ADS/JAO.ALMA\#2017.1.01784.S.\\ ALMA is a partnership of ESO (representing its member states), NSF (USA) and NINS (Japan), together with NRC (Canada), MOST and ASIAA (Taiwan), and KASI (Republic of Korea), in cooperation with the Republic of Chile. The Joint ALMA Observatory is operated by ESO, AUI/NRAO and NAOJ. The National Radio Astronomy Observatory is a facility of the National Science Foundation operated under cooperative agreement by Associated Universities, Inc. NK acknowledges support from the Department of Science and Technology via a J.~C.~Bose Fellowship (JCB/2023/000030), and from the Department of Atomic Energy, under project 12-R\&D-TFR-5.02-0700. This material is based upon work supported by the National Science Foundation under grant No. 2107989.
\end{acknowledgments}

\facilities{ALMA}
\software{ALMA pipeline \citep{Hunter2023}, Astropy \citep{Astropy2013,Astropy2018}, CASA \citep{CASA2022}, LineSeeker \citep{Gonzalez-Lopez2017}, Matplotlib \citep{Hunter2007}, Numpy \citep{Harris2020}, SoFiA \citep{Serra2015,Westmeier2021}, Qubefit \citep{Neeleman2021}}

\appendix

\section{Results from the automated search for line emission}
\label{sec:autoline_det}

\begin{figure*}[!tbh]
\includegraphics[width=\textwidth]{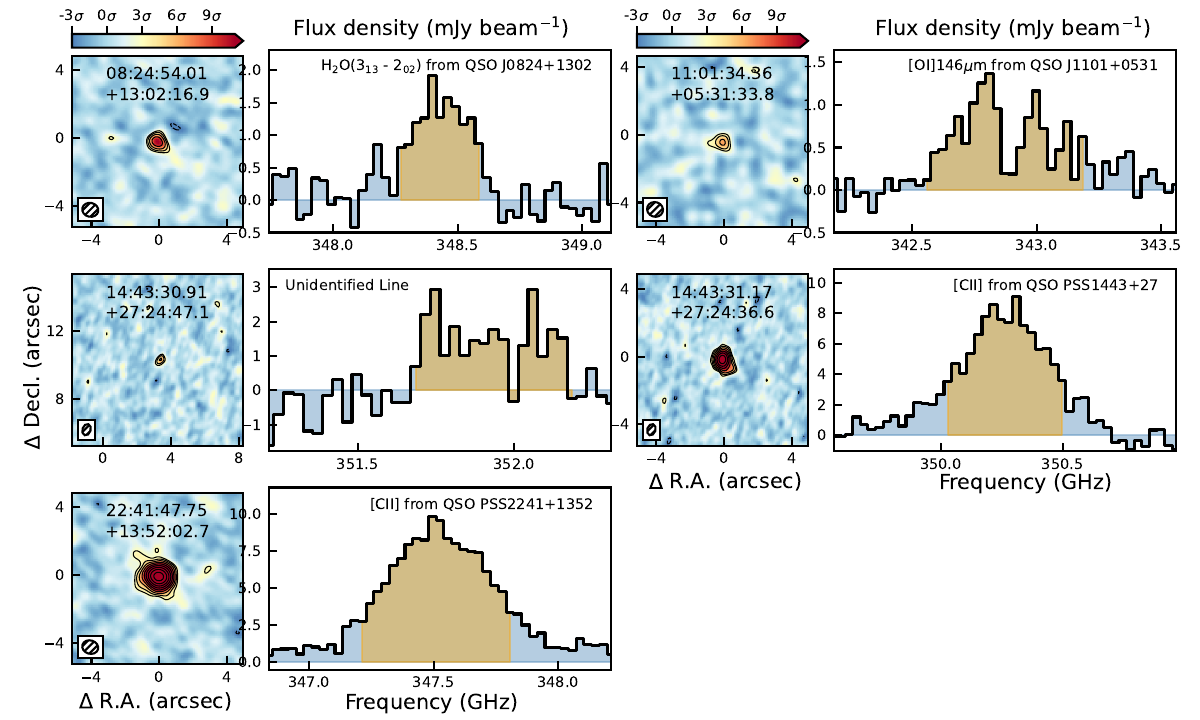}
\caption{Emission lines detected at $\geq 5.7\sigma$ significance in the full data set using the LineSeeker search algorithm that are either associated with the QSO or unidentified. The first and third columns show the velocity-integrated flux density maps for the line detections, integrated over a velocity range that is $1.2\times$ the FWHM of the emission line. In these panels, the x- and y-axes give the position relative to the pointing center of the observations, which in all cases corresponds to the optical position of the QSO. The contours in each panel start at 3$\sigma$ and increase by powers of $\sqrt{2}$. The second and fourth columns show the spectra (plotted at a velocity resolution of $\approx$~27~\kms) of either the central pixel of the line emission, or a $1\farcs0$-radius region centered at the peak pixel, depending on whether or not the line emission is spatially resolved (see main text for discussion). The orange region highlights the velocity width used for the integrated flux density map. The five line emitters found by LineSeeker and identified as \CII\ emitters at the DLA redshifts are shown in Figures~\ref{fig:dlagal} and \ref{fig:absems}.
\label{fig:lineauto}}
\end{figure*}

\begin{deluxetable*}{lrlllrrl}
\tablecaption{FIR properties of the line detections from the automated search
\label{tab:line}}
\tablehead{
\colhead{R.A.} &
\colhead{Decl.} &
\colhead{$\nu$} &
\colhead{FWHM} &
\colhead{$\int S_\nu dv$} &
\colhead{S/N$_{\rm LS}$} &
\colhead{S/N$_{\rm opt}$} &
\colhead{Line Identification}\\
\colhead{(J2000)} &
\colhead{(J2000)} &
\colhead{(GHz)} &
\colhead{(\kms)} &
\colhead{(Jy \kms)} &
\colhead{} &
\colhead{} &
\colhead{}
}
\startdata
08:17:40.87&+13:51:38.1&361.299 $\pm$ 0.006$^{\rm a}$&431 $\pm$  23$^{\rm a}$&4.18 $\pm$ 0.19$^{\rm a}$&23.3&42.1&\CII\ from DLA0817g1\\
08:24:54.01&+13:02:16.9&348.413 $\pm$ 0.011&239 $\pm$  21&0.40 $\pm$ 0.03&11.3&15.9&H$_2$O(3$_{13}$ - 2$_{02}$) from QSO J0824+1302\\
08:34:29.71&+21:40:23.2&352.637 $\pm$ 0.015&209 $\pm$  31&0.24 $\pm$ 0.03&7.3&9.9&\CII\ from DLA0834g1\\
10:54:45.66&+16:33:34.6&370.131 $\pm$ 0.014$^{\rm a}$&233 $\pm$  32$^{\rm a}$&2.36 $\pm$ 0.25$^{\rm a}$&10.7&14.7&\CII\ from DLA1054g1\\
11:01:34.36&+05:31:33.8&342.912 $\pm$ 0.012&464 $\pm$  53&0.40 $\pm$ 0.05&7.0&10.3&[\ion{O}{1}]146$\mu$m from QSO J1101+0531\\
14:43:30.91&+27:24:47.1&351.952 $\pm$ 0.022&359 $\pm$  69&0.63 $\pm$ 0.11&6.1&7.9&Unidentified Line\\
14:43:31.17&+27:24:36.6&350.259 $\pm$ 0.006$^{\rm a}$&356 $\pm$  13$^{\rm a}$&3.03 $\pm$ 0.09$^{\rm a}$&32.3&56.5&\CII\ from QSO PSS1443+27\\
14:43:31.29&+27:24:38.1&363.815 $\pm$ 0.031$^{\rm a}$&288 $\pm$  54$^{\rm a}$&0.87 $\pm$ 0.13$^{\rm a}$&8.6&12.5&\CII\ from DLA1443g1\\
16:26:25.86&+27:51:35.1&357.781 $\pm$ 0.014$^{\rm a}$&429 $\pm$  42$^{\rm a}$&2.51 $\pm$ 0.21$^{\rm a}$&13.8&16.8&\CII\ from DLA1626g1\\
22:41:47.75&+13:52:02.7&347.525 $\pm$ 0.004$^{\rm a}$&435 $\pm$  10$^{\rm a}$&4.01 $\pm$ 0.07$^{\rm a}$&62.2&88.1&\CII\ from QSO PSS2241+1352
\enddata
\tablecomments{Quantities marked with an $^{\rm a}$ are derived from a circular region with a radius of 1$''$ centered on the peak emission, because the emission is spatially resolved in the observations. For the other quantities,  the peak pixel is used.}
\end{deluxetable*}

In Table \ref{tab:line} we list all of the line emitters that are detected at a S/N $\ge 5.7$ within the ALMA primary beam (defined as the area where the primary beam response is $\geq 0.2$ times the central response). The table shows the J2000 coordinates of each emission feature, as well as the S/N determination from LineSeeker, S/N$_{\rm LS}$. The flux-weighted central frequency, $\nu$, and the integrated flux, $\int$$S_\nu dv$, are derived from the spectrum through either the peak pixel of the emission or from a $1\farcs0$ radius region centered at the peak pixel, depending on whether or not the emission is spatially resolved in the spectral cube. We define a source as spatially resolved if the integrated flux density over the extended region is at least $1\sigma$ higher than the integrated flux density measured from the spectrum through the peak pixel, where $\sigma$ is the rms error on the integrated flux density. The FWHM of the emission line is estimated from a Gaussian fit to the spectrum. Finally the optimum signal-to-noise ratio, S/N$_{\rm opt}$, is determined by integrating the flux density (and the corresponding uncertainty) over $1.2 \times$ the FWHM of the emission line. This velocity range maximizes the S/N for a Gaussian emission profile \citep[see, e.g.,][]{Decarli2018,Neeleman2021}. We note that all of our emission lines (except for the \CII\ emission line from DLA0817+1351g1) are all well-approximated by a Gaussian emission profile, justifying this assumption. 

Out of the ten detected line emitters with $\geq 5.7\sigma$ significance, we identify four of the lines as being associated with the background QSO (see Appendix~\ref{sec:qsoanalysis}), and five of the lines as arising from galaxies associated with the target DLAs (see Section~\ref{sec:firfromdla}). The last (and weakest) of the ten lines, in the field of PSS1443+27, is the only line emitter that has not been identified. However, its frequency suggests that it is not an emission line from a galaxy at either the redshift of the QSO or the DLA. This field has HST imaging available \citep{Kaur2021}, but we did not detect any near-infrared emission at the position of the line emission (down to a $2\sigma$ limiting magnitude of $\rm m_{AB} = 25.3$~arcsec$^{-2}$. The four lines associated with the background QSOs and the unidentified line are shown in Figure~\ref{fig:lineauto}.

\section{Results from the visual search for line emission}
\label{sec:manline_det}

\begin{deluxetable*}{lrlllr}
\tablecaption{FIR properties of the line emission candidates from the visual search
\label{tab:lineman}}
\tablehead{
\colhead{R.A.} &
\colhead{Decl.} &
\colhead{$\nu$} &
\colhead{FWHM} &
\colhead{$\int S_\nu dv$} &
\colhead{S/N$_{\rm opt}$}\\
\colhead{(J2000)} &
\colhead{(J2000)} &
\colhead{(GHz)} &
\colhead{(\kms)} &
\colhead{(Jy \kms)} &
\colhead{}
}
\startdata
08:17:40.87&+13:51:38.1&361.299 $\pm$ 0.006$^{\rm a}$&431 $\pm$  23$^{\rm a}$&4.18 $\pm$ 0.19$^{\rm a}$&42.1\\
08:34:29.71&+21:40:23.2&352.637 $\pm$ 0.015&209 $\pm$  31&0.24 $\pm$ 0.03&9.9\\
10:54:45.66&+16:33:34.6&370.131 $\pm$ 0.014$^{\rm a}$&233 $\pm$  32$^{\rm a}$&2.36 $\pm$ 0.25$^{\rm a}$&14.7\\
10:54:45.93&+16:33:35.9&370.101 $\pm$ 0.020&327 $\pm$  73&1.41 $\pm$ 0.25&6.7\\
14:43:31.29&+27:24:38.1&363.815 $\pm$ 0.031$^{\rm a}$&288 $\pm$  54$^{\rm a}$&0.87 $\pm$ 0.13$^{\rm a}$&12.5\\
14:43:31.41&+27:24:38.8&363.571 $\pm$ 0.019&228 $\pm$  43&0.31 $\pm$ 0.05&7.8\\
16:26:25.86&+27:51:35.1&357.781 $\pm$ 0.014$^{\rm a}$&429 $\pm$  42$^{\rm a}$&2.51 $\pm$ 0.21$^{\rm a}$&16.8\\
\tableline
08:34:28.80&+21:40:28.5&348.231 $\pm$ 0.020&187 $\pm$  72&0.25 $\pm$ 0.07&4.0\\
08:34:28.91&+21:40:21.1&352.771 $\pm$ 0.026$^{\rm a}$& 58 $\pm$  29$^{\rm a}$&0.18 $\pm$ 0.07$^{\rm a}$&4.7\\
09:53:55.64&$-$05:04:29.3&365.486 $\pm$ 0.013& 67 $\pm$  17&0.46 $\pm$ 0.09&5.6\\
11:00:44.65&+11:22:38.4&352.190 $\pm$ 0.004& 31 $\pm$   9&0.08 $\pm$ 0.02&5.1\\
11:01:34.33&+05:31:37.9&355.639 $\pm$ 0.016& 89 $\pm$  29&0.11 $\pm$ 0.03&4.9\\
12:05:22.83&$-$07:42:26.2&353.279 $\pm$ 0.007& 42 $\pm$  12&0.14 $\pm$ 0.03&5.1\\
12:05:23.25&$-$07:42:38.6&353.343 $\pm$ 0.003& 31 $\pm$   8&0.11 $\pm$ 0.02&5.4\\
12:05:23.65&$-$07:42:36.8&353.050 $\pm$ 0.010& 50 $\pm$  13&0.20 $\pm$ 0.05&5.4\\
14:43:31.55&+27:24:28.7&364.171 $\pm$ 0.017&152 $\pm$  28&0.57 $\pm$ 0.09&7.9\\
22:41:47.76&+13:52:04.9&359.705 $\pm$ 0.021& 99 $\pm$  33&0.12 $\pm$ 0.03&4.4
\enddata
\tablecomments{Quantities marked with an $^{\rm a}$ are derived from a circular region with a radius of 1$''$ centered on the peak emission, because the emission is spatially resolved in the observations. For the other quantities the peak pixel is used. Emission features above the horizontal line in the table are treated as detections, whereas the features below this line are considered to be line candidates. S/N$_{\rm opt}$ denotes the higher optimum S/N in the two independent analyses.}
\end{deluxetable*}

\begin{figure*}[!tbh]
\includegraphics[width=\textwidth]{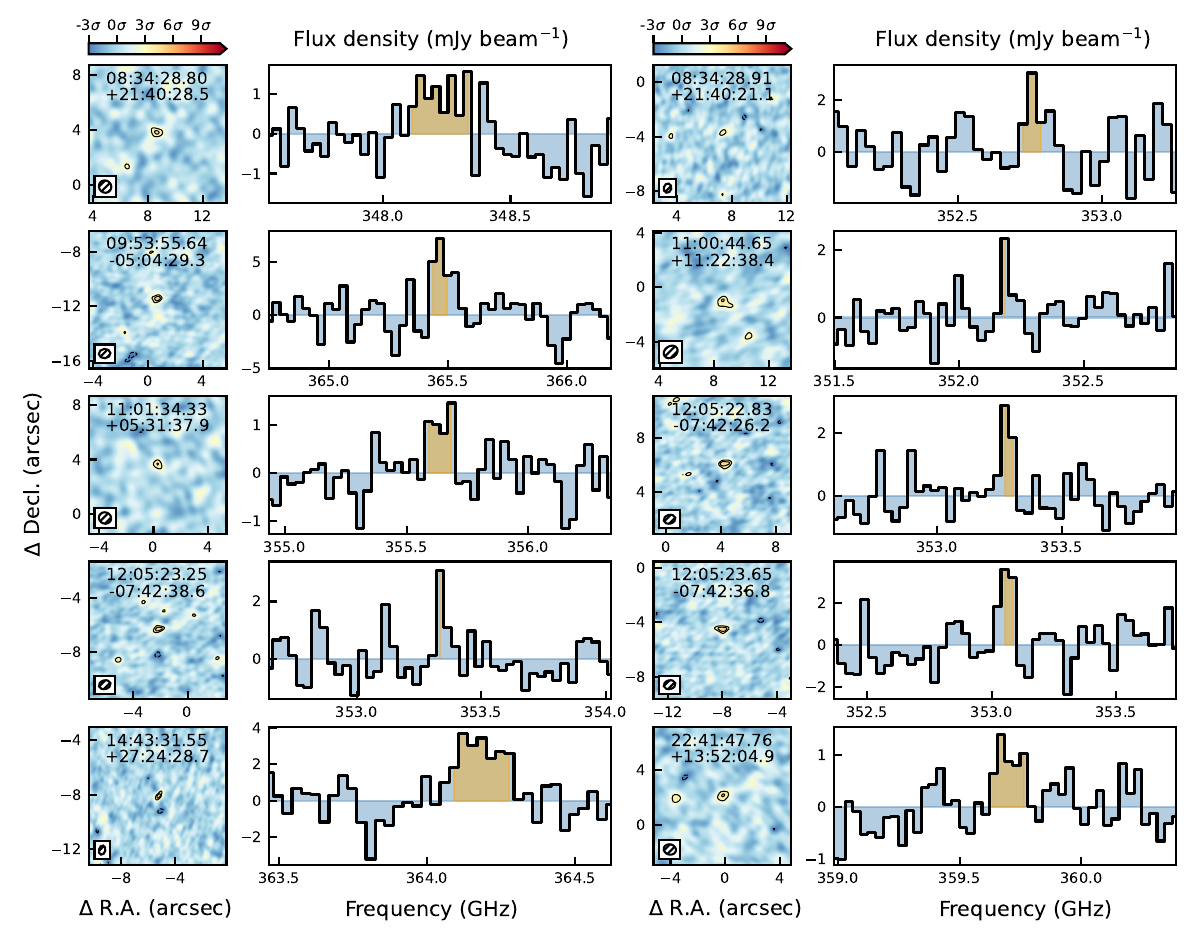}
\caption{Emission line candidates lying within the search area of the target DLAs that were identified via visual searches. The seven line emitters that were identified as \CII\ emission from DLA galaxies (Figures~\ref{fig:dlagal} and \ref{fig:absems}) are not shown here. The first and third columns show the integrated flux density maps for the line detections, integrated over a region that is $1.2\times$ the FWHM of the emission feature. In these panels, the x- and y-axes give the position relative to the pointing center of the observations, which in all cases corresponds to the optical position of the QSO. The contours in each panel start at 3$\sigma$ and increase by powers of $\sqrt{2}$. The second and fourth columns show the spectra (plotted at a velocity resolution of $\approx$~27~\kms) of either the central pixel of the emission or a $1\farcs0$-radius region centered at the peak pixel, depending on whether or not the emission is spatially resolved (see main text for discussion). The orange regions in the spectra indicates the velocity extents used for the integrated flux density map.
\label{fig:lineman}}
\end{figure*}

The visual line search resulted in a total of 17 emission line candidates detected in two independently reduced data sets, with  S/N $\geq 5$. Details of these 17 emission line candidates are given in Table~\ref{tab:lineman}. Seven candidates were detected at $\geq 5\sigma$ significance in both independent analyses, of which five were also found by the automated search. The remaining two line candidates were previously reported in \citet{Neeleman2019} and \citet{Kaur2021} for 14:43:31.29 $+$27:24:38.1 and 10:54:45.93 $+$16:33:35.9, respectively. All these line candidates are well within the primary beam of the ALMA observations, have relatively wide emission and have high S/N ratios. This indicates that these features are unlikely to be spurious. These seven line emission candidates constitute our sample of \CII\ emitters associated with the target DLAs; they are shown in Figures~\ref{fig:dlagal} and \ref{fig:absems}. The remaining 10 emission line candidates were detected in one of the independent analyses at a S/N~$\geq  5$, but not in the other. Figure \ref{fig:lineman} shows these ten emission line candidates. Six of the candidates have narrow emission lines, which can arise from noise fluctuations in the data cube and hence may be spurious. Using these updated definitions, we classify these relatively narrow emission lines as line candidates, and exclude them from the statistical analysis performed in this paper. The emission line features at 11:01:34.33 $+$05:31:37.9, 08:34:28.80 $+$21:40.28.5 and 22:41:47.76 $+$13:52:04.9 are all relatively wide, but the S/N on each feature was $< 5$ in one of the two independent analysis; we hence chose to classify these features as candidate detections, and excluded them from the statistical analysis. Note that the first of these features was reported as a detection in \citet{Neeleman2019}. Finally, the line candidate at 14:43:31.55 $+$27:24:28.7 is both bright and wide, but it is located towards the edge of the spectral range and at the edge of the primary beam. This decreases the fidelity of the line candidate because of larger uncertainties in the continuum subtraction and the response of the primary beam, respectively. We therefore conservatively classify this too as a line candidate. 

\section{Results from the search for far-infrared continuum emission}
\label{sec:contdet}

The sources identified in the continuum images of our DLA fields are listed in Table~\ref{tab:contdet}, with the continuum images shown in Figure~\ref{fig:cont}. We detect continuum emission from the QSO in all fields, with flux densities in the range $0.46 \pm 0.04$~mJy (J0817+1351; S/N~$\approx 12$) and $16.04 \pm 0.07$~mJy (BR1202$-$07; S/N~$\approx 230$). There are 14 additional continuum sources detected at $> 4.4\sigma$ significance within our fields. Two of these coincide spatially with the \CII\ line emission from the DLA galaxies, DLA0817g1 and DLA1626g1, and we assign the continuum emission to these galaxies. In addition, for the well-known field surrounding BR1202$-$07, we recover the continuum sources detected in previous studies \citep[e.g.][]{Wagg2012, Drake2020}. Of the remaining continuum sources, a particularly interesting identification is the bright ($8.00 \pm 0.04$ mJy) source adjacent to QSO~J1626+2751. The on-sky separation between the QSO and the continuum source is about $1\farcs1$, suggesting that this is an interacting pair of galaxies, one a quasar and the other a sub-millimeter galaxy (SMG). This is similar to BR1202$-$07, albeit with a much smaller angular separation. However, higher spectral and spatial resolution observations are needed to determine the redshift of the SMG, and thus confirm that the galaxy is at the same redshift as the quasar. We further detect two continuum sources in the fields of both QSO~J0834+2140 and QSO~J1607+1604. Their proximity to the quasar suggests that these quasars reside in galaxy groups, although the redshift of the continuum sources needs to be determined to confirm this hypothesis. 

\begin{deluxetable}{llrcc}
\tablecaption{FIR properties of the continuum detections
\label{tab:contdet}}
\tablehead{
\colhead{Name} &
\colhead{R.A.} &
\colhead{Decl.} &
\colhead{$\nu$} &
\colhead{$S_{\nu}$}\\
\colhead{} &
\colhead{(J2000)} &
\colhead{J(2000)} &
\colhead{(GHz)} &
\colhead{(mJy)}
}
\startdata
J0307$-$4945$^{\rm a}$&03:07:22.89&$-$49:45:48.3&341&1.83 $\pm$ 0.03\\
J0307$-$4945C1&03:07:23.52&$-$49:45:48.4&341&0.31 $\pm$ 0.02\\
J0817+1351&08:17:40.53&+13:51:34.5&356&0.46 $\pm$ 0.04\\
J0817+1351C1$^{\rm a}$&08:17:40.86&+13:51:38.2&356&1.18 $\pm$ 0.10\\
J0824+1302$^{\rm a}$&08:24:54.01&+13:02:17.0&342&13.11 $\pm$ 0.05\\
J0834+2140$^{\rm a}$&08:34:29.44&+21:40:24.7&343&2.44 $\pm$ 0.04\\
J0834+2140C1$^{\rm a}$&08:34:29.31&+21:40:24.4&343&0.25 $\pm$ 0.04\\
J0834+2140C2&08:34:29.62&+21:40:25.7&345&0.10 $\pm$ 0.02\\
BR0951$-$04$^{\rm a}$&09:53:55.75&$-$05:04:19.0&360&1.18 $\pm$ 0.05\\
J1054+1633$^{\rm a}$&10:54:45.42&+16:33:37.6&363&2.23 $\pm$ 0.10\\
J1054+1633C1&10:54:45.53&+16:33:38.4&363&0.27 $\pm$ 0.06\\
J1100+1122$^{\rm a}$&11:00:45.23&+11:22:39.1&345&0.52 $\pm$ 0.03\\
J1101+0531$^{\rm a}$&11:01:34.36&+05:31:33.8&349&8.77 $\pm$ 0.05\\
J1132+1209&11:32:46.50&+12:09:01.7&348&2.46 $\pm$ 0.02\\
BR1202$-$07$^{\rm a}$&12:05:23.13&$-$07:42:32.8&348&16.04 $\pm$ 0.07\\
BR1202$-$07C1$^{\rm a}$&12:05:22.98&$-$07:42:29.7&348&18.86 $\pm$ 0.08\\
BR1202$-$07C2$^{\rm a}$&12:05:23.06&$-$07:42:34.5&348&1.94 $\pm$ 0.07\\
BR1202$-$07C3$^{\rm a}$&12:05:23.46&$-$07:42:32.2&348&0.47 $\pm$ 0.09\\
PSS1443+27$^{\rm a}$&14:43:31.17&+27:24:36.8&357&1.59 $\pm$ 0.08\\
PSS1443+27C1&14:43:30.78&+27:24:32.8&357&0.19 $\pm$ 0.04\\
J1607+1604$^{\rm a}$&16:07:34.22&+16:04:17.4&342&2.96 $\pm$ 0.04\\
J1607+1604C1$^{\rm a}$&16:07:34.30&+16:04:14.7&342&0.37 $\pm$ 0.04\\
J1607+1604C2$^{\rm a}$&16:07:34.45&+16:04:16.3&342&0.31 $\pm$ 0.04\\
J1626+2751$^{\rm a}$&16:26:26.51&+27:51:32.4&353&4.89 $\pm$ 0.04\\
J1626+2751C1$^{\rm a}$&16:26:26.55&+27:51:33.5&353&8.00 $\pm$ 0.04\\
J1626+2751C2&16:26:25.86&+27:51:35.3&353&0.30 $\pm$ 0.06\\
PSS2241+1352$^{\rm a}$&22:41:47.75&+13:52:02.8&355&2.89 $\pm$ 0.06
\enddata
\tablecomments{Lines marked with an $^{\rm a}$ are derived from a circular region with a radius of 1$''$ centered on the peak emission, because the emission is spatially resolved in the observations. For the other quantities the peak pixel is used.}
\end{deluxetable}

\section{FIR emission from the QSOs}
\label{sec:qsoanalysis}

\begin{figure*}[!tbh]
\includegraphics[width=\textwidth]{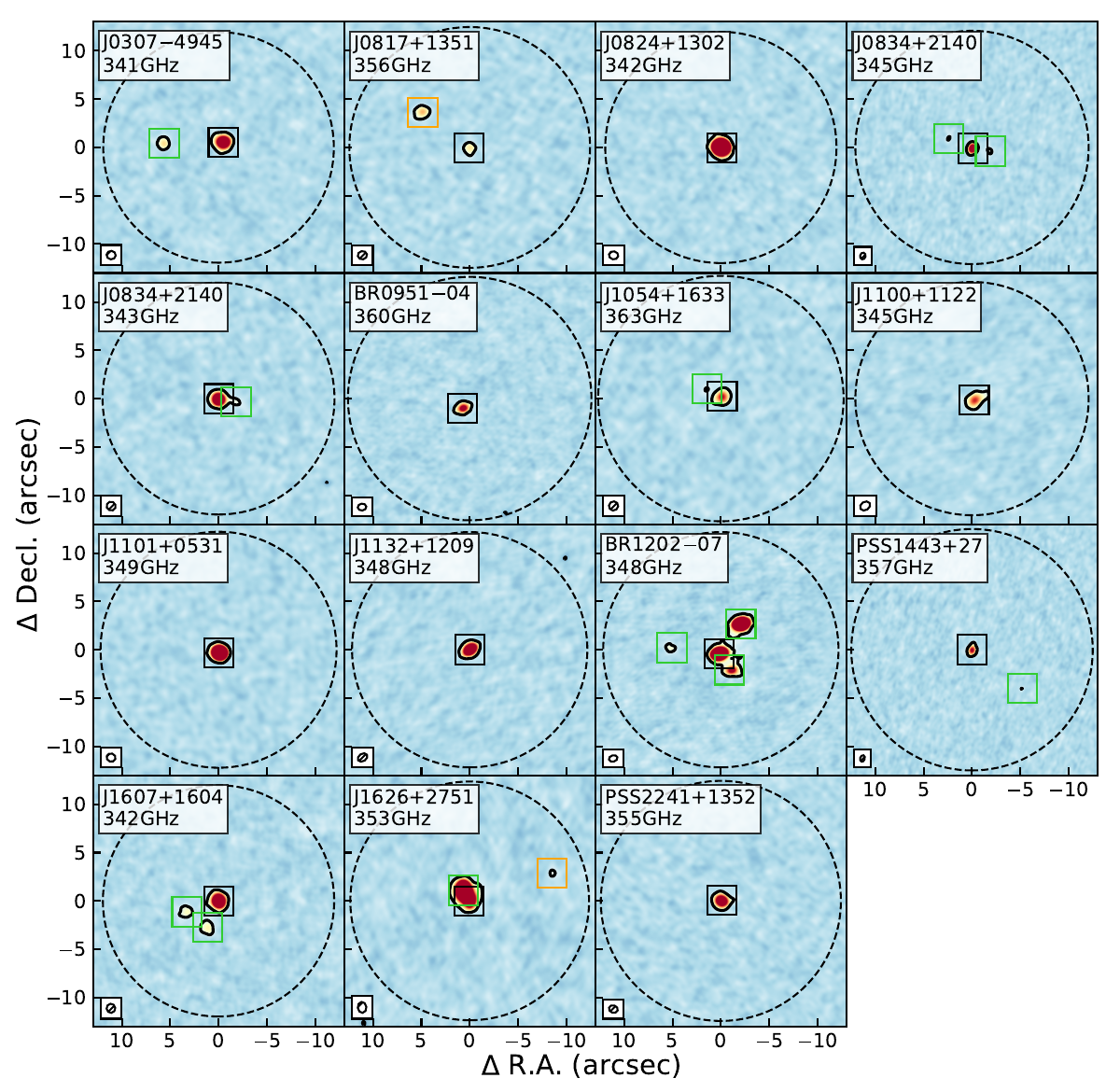}
\caption{ALMA continuum images of the 15 DLA fields. The axes are relative to the QSO position. All continuum sources with S/N~$\geq 4.4$ are boxed and the $4.4\sigma$ contour is shown. Black boxes correspond to emission from the QSO, green boxes are companion sources, and orange boxes denote continuum emission associated with DLA galaxies. The primary beam of the observations (defined as the area where the primary beam response is greater than 0.2) is shown by the dashed circle, whereas the synthesized beam is shown in the bottom left inset. The field surrounding QSO J0834+2140 was observed at two different spectral and spatial resolutions. In both observations, the quasar and the brightest companion (J0834+2140C1) were detected, whereas the second source remained undetected (S/N~$< 4.4$) in the lower-resolution, lower-sensitivity image. The field of QSO J1626+2751, which has two DLAs, has only one continuum image, as the same setup covered the redshifted \CII\ frequencies of both DLAs.
\label{fig:cont}}
\end{figure*}

In our sample, we serendipitously detected four emission lines spatially coincident with the background optical QSO. We associate these with the QSO, and use the QSO redshift (estimated from either rest-frame ultra-violet emission lines such as \lya\ or \ion{C}{4}, or the onset of the \lya\ forest), to identify the spectral transition that gives rise to the observed emission line. In all four cases, we identified the most-likely transition responsible for the emission; these are listed in Table~\ref{tab:line}. These FIR emission lines provide updated --- and much more reliable --- redshifts for the QSOs: $z=5.2114 \pm 0.0002$, $z=5.0082 \pm 0.0002$,  $z=4.4262 \pm 0.0002$ and  $z=4.4688 \pm 0.0001$ for the QSOs J0824$+$1302, J1101$+$0531, PSS14443$+$27, and PSS2241$+$2751, respectively. All these redshifts are slightly higher (on average, by $\approx 2000$~\kms) than the redshift estimates from the rest-frame UV lines. This is consistent with the observed velocity offsets between UV and FIR emission lines in other high redshift quasars \citep[e.g.][]{Meyer2019,Schindler2020}.

In Figure~\ref{fig:cont}, we show the continuum images of the QSO fields. In all cases, we detect the underlying continuum from the 14 QSOs. In addition, we detect continuum emission from 14 other sources, as discussed in Appendix~\ref{sec:contdet}. We can provide a rough estimate of the clustering of galaxies around these quasars from our continuum detections, comparing the number of sources to those identified in blank fields \citep[e.g.,][]{Fujimoto2016}. Following the approach of \citet{Champagne2018}, we expect a total of $N_{\rm exp} = 7 \pm 3$ continuum sources with flux density $ \geq 0.1$~mJy within the primary beam of our 14 independent fields\footnote{Our observations of the two ``double-DLA pairs'' toward J0834+2140 and J1626+2751 cover the same continuum fields.}. Excluding the two continuum sources that arise from DLA galaxies (Section~\ref{sec:firfromdla}), we detected $N_{\rm gal} = 12 \pm 3$ continuum sources in our 14 fields. This results in an overdensity factor of $\delta_{\rm gal} = (N_{\rm gal} - N_{\rm exp}) / N_{\rm exp} = 0.71 \pm 0.85$. This estimate of the average overdensity in the 14 fields is consistent with no overdensity, despite the presence of the known overdense region surrounding  BR1202$-$07 \citep{Wagg2012, Drake2020} in our sample. The lack of a systematic overdensity in dust continuum sources in the fields of high-redshift quasars is consistent with previous results \citep{Champagne2018}.

\bibliography{ms}
\bibliographystyle{aasjournal}

\end{document}